# A Data-Driven Based Concurrent Coupling Approach for Cryogenic Propellant Tank Long-term Pressure Control Predictions

Qiyun Cheng, Huihua Yang, Shanbin Shi, and Wei Ji*

Department of Mechanical, Aerospace, and Nuclear Engineering, Rensselaer Polytechnic Institute, 110 8th Street, Troy, NY 12018

**Abstract**

The design and optimization of cryogenic propellant storage tanks for NASA's future space missions require fast and accurate predictions of long-term fluid behaviors. Computational fluid dynamics (CFD) techniques are high-fidelity but computationally prohibitive. Coarse mesh nodal techniques are fast but heavily rely on empirical correlations to capture prominent three-dimensional phenomena. A data-driven based concurrent coupling (DCC) approach has been developed to integrate CFD with nodal techniques for efficient and accurate analysis. This concurrent coupling scheme generates case-specific correlations on the fly through a short period of co-solving CFD and nodal simulations, followed by a long-period nodal simulation with CFD-corrected solutions. This paper presents the approach development, stability analysis, and efficiency demonstration, specifically for modeling two-phase cryogenic propellant tank self-pressurization and periodic mixing phenomena. Linear regression is employed to derive the data-driven correlations. The self-pressurization experiments of Multipurpose Hydrogen Test Bed experiments and K-Site tank are used to validate the approach. The DCC approach accurately predicts temperature stratifications while reducing computational time by as much as 70% compared to pure CFD simulations. Additionally, the DCC approach mitigates the risks of numerical instability and correlation loss inherent in current domain decomposition or overlapping-based coupling methods, making it a flexible and user-friendly approach for integrated CFD and nodal analysis of cryogenic tank behaviors.

## 1. Introduction

NASA's future Moon and Mars missions require extended storage of cryogenic fluids for propulsion and life support. However, heat leakage from background radiation and structural contacts will lead to thermal stratification and pressurization of cryogenic storage tanks. To address this issue, minimizing heat leakage in tank design and employing appropriate depressurization procedures are crucial, identified by NASA as having the greatest potential for cost savings [1]. Experiments have been conducted in both ground-based [2-5] and microgravity conditions [6,7], to investigate thermal stratification phenomena and pressure control approaches. However, the inherent uncertainty of experiments makes it difficult to accurately identify experimental conditions and other sensible design parameters [8], lacking sufficient details for validating tank designs. Additionally, the high cost of space experiments limits microgravity experiments, particularly for large-size cryogenic tanks. Thus, highly accurate and efficient numerical simulation tools are pivotal in reducing costs and enhancing the designs of cryogenic tanks for long-term space missions [9].

* Corresponding author: jiw2@rpi.edu

Among the existing numerical tools, lumped parameter (nodal) and computational fluid dynamics (CFD) codes are predominant approaches for cryogenic tank simulations. Nodal codes, such as SINDA/FLUINT [10], TankSIM [11], depict intricate thermal hydraulic systems as interconnected networks of zero-dimensional or one-dimensional components and are favored for system-level simulations. Due to the coarse mesh nature in 1D equations, it is rather challenging for nodal codes when 3D phenomena are prominent, such as thermal stratification [12] or jet mixing [13] in cryogenic tanks. Thus, empirical correlations are usually involved in nodal models to capture sub-grid physics like natural convection. Several groups incorporate the nodal codes with the empirical correlations and successfully improve the accuracy of the cryogenic tank simulations [14,15,16]. However, the applicable range of the empirical correlations must be well understood [17]. For more complex scenarios like sloshing, reliable and universal empirical correlations are still under development. Furthermore, extending existing empirical correlations to microgravity conditions raises concerns, while formulating new correlations suited for microgravity is difficult and expensive due to the lack of experimental data. In contrast, the increasing maturity and high-fidelity 3D capabilities of CFD codes have attracted growing interest. Because of the lesser simplifications and the utilization of the 3D governing equations, CFD simulations are expected to perform better than the nodal simulations for complex geometry and intricate 3D phenomenon [18-20]. Commercial CFD software such as Fluent [21] and Flow-3D [22] have exhibited remarkable performance in simulating cryogenic tanks for various NASA testing sites, such as MHTB [23], and ZBOT [24]. Open source CFD software, such as the OpenFOAM, is also demonstrated using small cryogenic tank experiments [25]. However, due to their substantial computational requirements, CFD simulations often demand weeks to yield results for events lasting only a few hours in real time. This computational overhead poses a significant limitation for parametric analysis and the iterative optimization for tank design and mission planning.

A promising methodology to tackle the difficulty is to perform an integrated analysis by coupling the CFD and nodal codes. This strategy capitalizes on the high-resolution fluid field capabilities of CFD codes and the fast execution of nodal codes [26]. Several coupling approaches have been developed and demonstrated for various thermos-fluid problems [27-34]. The most straightforward idea is the spatial domain decomposition coupling approach. This coupling approach partitions the computational domain into several sub-domains and uses either CFD or nodal code for each sub-domain. Sub-domains with complex fluid behaviors are solved by the CFD codes, and the solution data of different codes at the sub-domain interfaces is transferred as the boundary conditions to collectively solve the entire problem. The spatial domain decomposition coupling approach has been implemented with different combinations of CFD and nodal codes, such as Fluent-RELAP5 [27], and CFX-TRACE [28]. Algorithms to accelerate the iterative solving of different spatial domains in the spatial domain overlapping approach have also been developed [29, 30]. The algorithms are implemented using the Multiphysics Object-Oriented Simulation Environment (MOOSE) for conjugate heat transfer problems. However, the spatial domain decomposition coupling approach necessitates small time steps and/or enough code-to-code iterations within time steps to avoid numerical instabilities [31]. Another method is the spatial domain overlapping coupling approach, which applies the nodal code across the entire computational domain while utilizing the CFD code in an overlapping manner for a smaller portion of the domain. The CFD solutions refine empirical coefficients, such as the friction coefficient, in the nodal code simulation at each time step. Although this approach enhances numerical stability, it demands careful design to accurately adjust the 1D empirical correlation coefficients with the 3D CFD solutions. However, a notable limitation is the ignorance of the inherent relationships between different variables, which is identified by Zhang and Sanchez [35] as the primary defect of correcting parameters independently. These issues introduce challenges to implementing of the domain overlapping coupling approach. Coupling works based on the spatial domain overlapping approach are also reported with different codes, such as FLUENT- ATHLET [32] and STAR-CCM+-TRACE [33]. Besides directly coupling the two codes, the data assimilation approach is also discussed to use high fidelity or experiment data to continuously improve the 0D nodal simulations through the reinforcement learning process [34].

For cryogenic tank simulations, the spatial domain decomposition approach is more convenient since there is a clear interface between the vapor and liquid that can serve as the decomposition interface. Panzarella and Kassemi [36] implemented the spatial domain decomposition approach on a spherical cryogenic tank for self-pressurization simulations. They partitioned the tank at the vapor-liquid interface, treating the vapor region as a uniform lump, while simulating the liquid region using a modified version of the commercial Galerkin finite-element codes. Using that model, they investigate the effect of different heating configurations and the mass and energy transfer at the interface. However, the results of their coupling approach were solely compared with thermodynamic solutions, which assumed uniform temperature throughout the entire tank, without any comparison to experimental or high-fidelity solutions. Moradikazerouni et al. [37] used the spatial domain decomposition coupling approach and divided the cryogenic tank at the vapor-liquid interface. In their model, the vapor region is simulated using the nodal code while the liquid region is solved using the CFD. Furthermore, they simplified the CFD simulation by employing Sparse Identification of Nonlinear Dynamics (SINDy) [38] package to reduce the computation time. However, all the aforementioned coupling studies encountered challenges related to 1D to 3D data transfer errors at each coupling time step, and the substantial computational costs from the involved CFD simulations remained unresolved. In microgravity scenarios, the spherical interface will also introduce more challenges for the domain decomposition approach.

In the above discussion, it is evident that current coupling approaches still lack the efficiency and flexibility required for long-term simulations of cryogenic storage tanks. In recent years, data-driven based approaches gained attention in various fields. In cryogenic tank simulations, researchers try to use data-driven approaches to optimize the existing empirical correlations or develop new correlations from experimental data or numerical simulations. However, attempting to formulate universal correlations are always difficult since the applicable range and the generalization performance of these correlations are questionable. Therefore, in this work, a data-driven based approach is adapted with a concurrent coupling scheme, named as data-driven concurrent coupling (DCC), is developed for simulating cryogenic tank pressure control behaviors. Instead of simply adding extra terms to compensate the missing sub-grid 3D information from different effects, the DCC approach directly maps the governing equations of the nodal codes to those of the CFD codes to enhance the accuracy of long-term system-level simulations [39]. The DCC approach begins with an initial co-solving of the problem using both CFD and nodal codes over a small fraction of the entire simulation period. Then, using the CFD simulations as high-fidelity benchmarks, solutions from co-solving are used by data-driven approaches to obtain the problem-dependent empirical correlations for interested variables, such as lump temperature and pressure, by directly mapping solutions from CFD to nodal codes. After that, the nodal code runs exclusively with the data-driven correlations to continue the simulation for a more extended period, delivering solutions that are "CFD-level" accurate. Thus, the DCC approach is expected to substantially reduce computational costs for long-term simulations, making it comparable to nodal code simulations while eliminating the need for 1D to 3D (or 3D to 1D) data transfer within time steps.

The present paper is structured as follows: Section 2 briefly introduces the basics of nodal and CFD codes simulations. Section 3 formulates the algorithm of the data-driven based coupling approach. The linear regression extrapolation error of the method is also discussed. Section 4 demonstrates the applications of the DCC approach using a two-phase cryogenic tank model. A tank self-pressurization process with the thermal stratification phenomenon is first studied. Linear regression is used to map the nodal code energy equations into the corresponding volume averaged CFD energy equations for temperature predictions. Then a problem representing a cyclic pressurization-depressurization process in a cryogenic tank is solved using the DCC approach. The pressure and temperature evolutions in tanks from the DCC approach are compared against the CFD solutions to demonstrate the effectiveness of the proposed approach. Finally, the self-pressurization experiments of Multipurpose Hydrogen Test Bed (MHTB) experiments and K-Site tank are used to validate the DCC approach. Section 5 summarizes the performance and limitations of the DCC approach.

## 2. Modeling Approach Descriptions

This section provides an overview of the numerical models and methodologies employed in this study. Considering the growing maturity of CFD codes for simulating two-phase flow, the two-phase flow models utilized in both CFD and nodal codes are also discussed, which are then implemented to demonstrate the DCC approach's capability in handling complex fluid problems.

### 2.1. Computational Fluid Dynamics (CFD) Models

Reynolds Averaged Navier-Stokes (RANS) equations are used for the CFD simulations in this paper. Since the compressible fluid is used, the continuity equation is:

$$\frac{\partial \rho}{\partial t} + \nabla \cdot (\rho \vec{v}) = S_C \ , \tag{1}$$

where $\rho(\vec{r}, t)$ is the density of the fluid, $\vec{v}(\vec{r}, t)$ is the velocity field, and $S_c(\vec{r}, t)$ is the source term. The momentum equation for Newtonian fluid is:

$$\frac{\partial(\rho \vec{v})}{\partial t} + \nabla \cdot (\rho \vec{v} \vec{v}) = -\nabla p + \nabla \cdot \left[\mu_{eff}(\nabla \vec{v} + \nabla \vec{v}^T)\right] + \rho \vec{g} + \overrightarrow{S_F} \ , \tag{2}$$

where $p(\vec{r}, t)$ is the pressure field, $\vec{g}$ is the gravitational acceleration, and $\overrightarrow{S_F}(\vec{r}, t)$ is the body force applied to the fluid. The effective viscosity, $\mu_{eff}(\vec{r}, t)$, includes the impact of the turbulent viscosity resulting from the Reynolds stress, which is determined by the selection of the turbulence model. In this work, a Laminar model is applied to slow-changing problems, such as self-pressurization, while the two equations Shear Stress Transport (SST) k-omega model [40] is selected for fast-changing events like venting and mixing. The energy equation expressed in terms of internal energy is:

$$\frac{\partial}{\partial t}(\rho u) + \nabla \cdot (\rho u \vec{v}) = -\nabla \cdot \vec{q}'' - p \nabla \cdot \vec{v} + \phi + S_E \ , \tag{3}$$

where $u(\vec{r}, t)$ is the specific internal energy, $\phi(\vec{r}, t)$ denotes the viscous dissipation, $\vec{q}''(\vec{r}, t)$ is the heat flux vector, and the $S_E(\vec{r}, t)$ is the volumetric energy source term.

To simulate the two-phase fluid, source terms on the right-hand side of the governing equations are modified to represent the phase interactions. Evaporation and condensation mass transfer terms are added into the $S_c(\vec{r}, t)$ in Equation 1:

$$S_C = \sum(\dot{m}_{pq} - \dot{m}_{qp}) \ , \tag{4}$$

where $\dot{m}(\vec{r}, t)$ represent the mass transfer rate, and subscripts $p$ and $q$ denote different phases. The Lee model [41] is used to determine the mass transfer rate for the phase change.

In the momentum equation, the body force term $\overrightarrow{S_F}(\vec{r}, t)$ takes into account the accumulated effects of the surface tension force. The Continuum Surface Force (CSF) model [42] is used for the surface tension forces at the liquid-vapor interface with the polynomial surface tension coefficients. Thus, surface tension forces at the interface are transformed into a body force term and is added into the $\overrightarrow{S_F}(\vec{r}, t)$.

The source term $S_E(\vec{r}, t)$ in Equation 3 is modified to account for the interface energy transfer due to the mass transfer between the two phases. Taking phase $p$ as an example, the corresponding source term $H_p = (\dot{m}_{qp} - \dot{m}_{pq})h_{qp}$ is added to $S_E(\vec{r}, t)$, where $h_{qp}$ is the latent heat of the fluid at saturation conditions. This term represents the energy change caused by phase change processes such as evaporation or condensation, with the sign of $H_p$ determining whether energy is absorbed or released.

The location of the interface between the two phases needs to be tracked and the volume of fluid (VOF) method with the sharp interface scheme [43] is utilized. The VOF method assigns a volume fraction to each phase such that the sum of volume fractions in each mesh element/cell is unity. Therefore, a single set of governing equations with volume fraction weighted filed variables is solved. The tracking of the interface between the phases is accomplished by solving a continuity equation for the volume fraction [21]:

$$\frac{\partial}{\partial t}\left(\alpha_q \rho_q\right) + \nabla \cdot \left(\alpha_q \rho_q \overrightarrow{v_q}\right) = S_c \, , \tag{5}$$

where $\alpha(\vec{r}, t)$ represents the volume fraction. More detailed derivations and successful demonstrations of the capabilities of the VOF model with sharp interface scheme can be referred to Refs. [20, 23, 24].

### 2.2. Nodal Code Model

The basic component of a nodal model is a zero-dimensional "lump", which represents a control volume containing mass and energy with a single equilibrium thermodynamic state. Therefore, perfect mixing of the fluid is inherently assumed inside a lump. Lumps follow the mass and energy conservation laws. For Lump $k$ with mass $M_k$, the mass conservation equation is:

$$\frac{dM_k}{dt} = \sum_j \dot{m}_{kj} n_{kj} \, , \tag{6}$$

where $\dot{\mathrm{m}}_{\mathrm{kj}}(t)$ is the absolute value of the mass flow rate between the lump $k$ and the $j^{th}$ neighboring lump. The direction is denoted by $n_{kj}$, which equals 1 or -1 based on the predefined positive flow direction. The governing equation for the internal energy $U_L$ of Lump $k$ is:

$$\frac{dU_{L,k}}{dt} = \sum_j \dot{m}_{kj} * h_{k/j} n_{kj} + \sum_j k_{eff,kj}\left(T_{L,j} - T_{L,k}\right) . \tag{7}$$

The first term on the right-hand side of the equation represents the energy exchange via mass transfer from adjacent lumps to the lump $k$, where $h_{k/j}(t)$ is the donor enthalpy in either lump $k$ or the $j^{th}$ neighboring lump. The second term represents energy exchange through heat transfer from adjacent lumps to lump $k$. Here, $T_L(t)$ denotes the lump temperature and $k_{eff,kj}$ is the effective conductivity between the target lump $k$ and the $j^{th}$ neighboring lump. The effective conductivity $k_{eff,kj}$ includes both conduction and convection heat transfer, which is usually computed from empirical correlations.

For two-phase fluid problems, a technique named twinned lump is utilized to simulate the fluid near the interface. A twinned lump divides a single lump into two control volumes with an imaginary interface, allowing different thermodynamic states in the two control volumes. The interface is assumed to be infinitely thin and at the saturation temperature of the current pressure. The mass transfer due to phase change is calculated based on the energy jump conditions at the interface:

$$\dot{m}_{evap} = \frac{H_v A_{if}(T_v - T_{sat}) - H_l A_{if}(T_{sat} - T_l)}{H_{fg}}, \quad (8)$$

where $\dot{m}_{evap}(t)$ is the evaporation mass flow rate and $A_{if}$ is the interface area. $T_v(t)$, $T_l(t)$, and $T_{sat}(t)$ are the temperatures of the vapor, liquid, and the saturated interface, respectively. The heat transfer coefficients of the vapor and the liquid are $H_v(t)$, and $H_l(t)$. $H_{fg}$ represents the latent heat of the phase change. To obtain heat transfer coefficients, empirical correlations, such as the Nusselt number correlation under the flat plate heat transfer condition [44], could be used as in Ref. [45].

Lumps are connected using "inertial tubes", which are one-dimensional pipes to account for velocity and pressure drops. The momentum equation for the Tube $k$ is given as follows:

$$\frac{d\dot{m}_k}{dt} = \frac{A_k}{L_k}\left(p_{up} - p_{down} + C_{1,k} + F_k\dot{m}_k|\dot{m}_k|^{C_{2,k}} + C_{3,k}\dot{m}_k^2 - \frac{K_k\dot{m}_k|\dot{m}_k|}{2\rho_{up}A_k^2}\right), \quad (9)$$

Where $A_k$ is the flow area and $L_k$ is the length of the tube. $p_{up}(t)$ and $p_{down}(t)$ are pressures of the lumps of the predefined upstream and downstream. $C_{1,k}$ represents the head coefficient due to body force, such as gravity or pump. $C_{3,k}$ is the recoverable loss coefficient caused by area and/or density change inside the tube. $K_k$ is also referred as the K factor, which comes from the entrances, exits or shape change (bends, tees, etc.) induced pressure loss. In cryogenic tank simulations, $C_{1,k}$, $C_{3,k}$ and $K_k$ are usually set as 0 [46]. $F_k$ is the friction coefficient, and $C_{2,k}$ is dependent on the flow regime, which is 0 for laminar flow and nearly 1 for turbulence flow. Empirical correlations, such as Darcy friction coefficients, are required to calculate $F_k$.

For a nodal model, a fluid system is first divided into distinct control volumes and each control volume is depicted using a lump. Then lumps are connected using tubes. Thus, the nodal model constructs a representation of the fluid system through an interconnected 0D/1D network. An illustration for a typical nodal model of a cylindrical two-phase tank is shown in Figure 1. Extra lumps representing boundary layers are usually used to better model the natural convection, while the volumes of the boundary lumps and the flowrates of the corresponding tubes have to be calculated using additional empirical correlations [45,46].

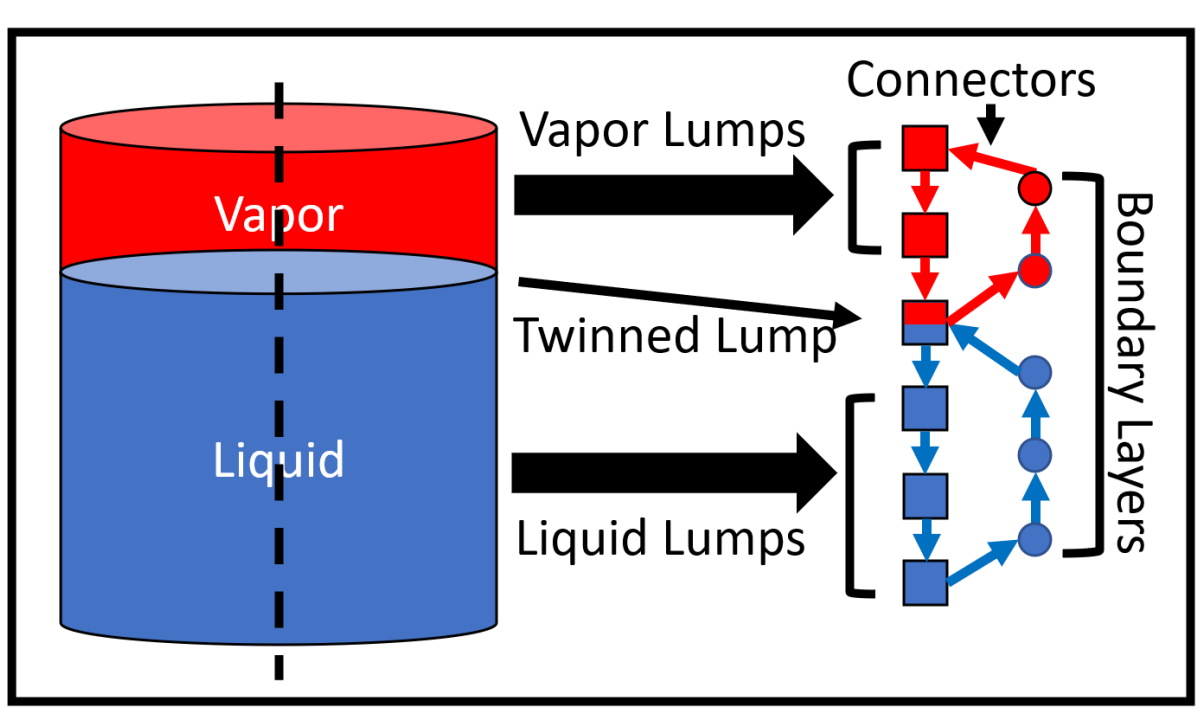

Figure 1. A nodal model for a two-phase cryogenic tank. Due to the symmetry, only half of a two-phase cryogenic tank is modeled.

In this work, reducing reliance on experimentally-derived empirical correlations is a primary motivation, especially for future micro-gravity simulations. Therefore, using empirical correlations is minimized in nodal models. Only the solid-like conduction model [10] is used for both vapor and liquid regions. This model treats each lump as a fluid bulk with 'solid-like' properties, using their respective thermal conductivities to compute heat transfer between the fluid bulks. Unlike relying on empirical correlations, the utilization of the solid-like conduction model represents a more universally applicable conduction process and remains unaffected by microgravity effects. Additionally, to illustrate the capability and versatility of the DCC approach, the nodal code model is intentionally designed with essential features while maintaining simplicity. No boundary layer lumps and natural convection empirical correlations are used. All heat transfer coefficients used in the nodal models, including the $k_{eff}$ in vapor and liquid regions, and the interface heat transfer coefficients, are calculated directly based on the corresponding thermal conductivities.

## 3. Data-driven Based Coupling Approach

In cryogenic tank problems, two typical scenarios often arise during pressure control processes. The first scenario involves long-term but slowly changing processes, such as the self-pressurization of cryogenic tanks. According to previous research by Panzarella and Kassemi [47], the self-pressurization is slow-changing and eventually (may take days or weeks) stabilizes at a steady pressurization rate during the long-term storage of cryogenic propellant. For the self-pressurization problems, thermal stratification within the tank makes nodal codes heavily reliant on empirical correlations to accurately simulate pressure and temperature behaviors. However, finding reliable empirical correlations is challenging, and such correlations may not even exist in microgravity conditions. The lengthy self-pressurization process also makes CFD simulations infeasible due to the computational cost. However, the slow-changing nature of these problems suggests that the differences between CFD and nodal simulations will also evolve slowly or remain stable over time. This argument motivates the DCC approach development. The second scenario involves periodic problems, which are commonly observed in cryogenic tanks due to pressurization-depressurization cycles. The depressurization process typically involves strong mixing and convection, which pose significant challenges for nodal simulations. Figure 2 presents a pressure evolution curve from a cryogenic tank experiment [48]. After self-pressurization, the tank undergoes periodic venting, displaying similar cycles with fluctuating starting and ending points. Similar patterns are also frequently observed in cryogenic tank sloshing problems [49]. Given the periodic nature of this scenario, the differences between CFD and nodal simulations can be expected to either remain consistent or change slowly from cycle to cycle. Therefore, both scenarios share a common challenge: long-term thermodynamic problems with significant 3D effects make nodal simulations inaccurate and CFD simulations computationally expensive. In the meanwhile, the differences between CFD and nodal simulations are expected to remain stable or change slowly over time.

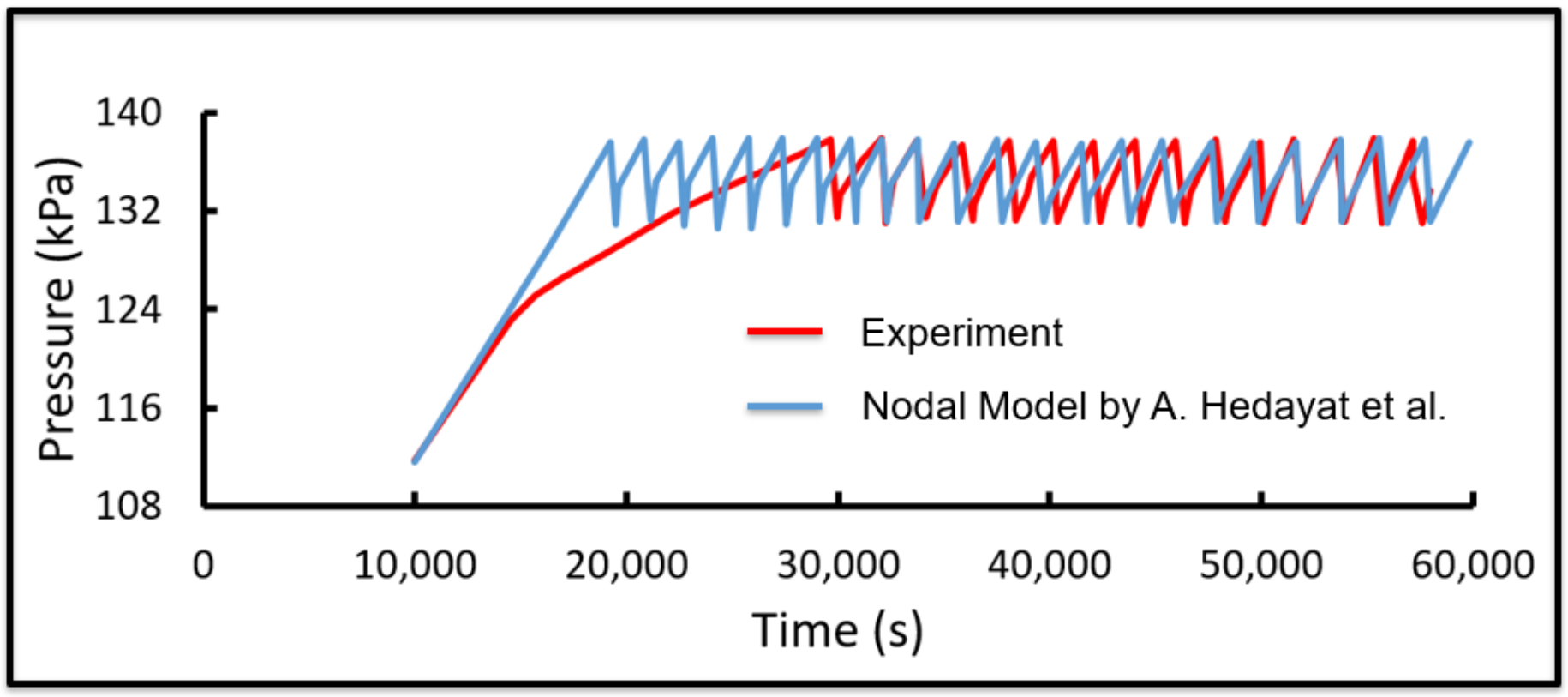

Figure 2. The periodic pressure evolution curve during the thermodynamic venting system (TVS) test in multipurpose hydrogen test bed (MHTB) [48].

For cryogenic tank simulations, nodal codes provide lump-based solutions, while CFD codes compute solutions at each fine mesh element within the same lump region, offering detailed fine-scale distributions of dynamic and thermal variables within the lump. Therefore, the differences between nodal and CFD codes arise from the averaging and simplifications inherent in the 1D governing equations of nodal codes. Since both CFD and nodal codes follow the same conservation laws, the lump-based nodal solutions are theoretically equivalent to the volume averaged CFD solutions when the boundary conditions are identical in each lump region. Thus, mapping correlations should exist that can convert the nodal codes' 1D governing equations into volume-averaged CFD equations at corresponding spatial resolutions, such as mapping lump temperatures to volume-averaged CFD temperatures in the same region. Section 3.1 will discuss more details about the comparison between the two sets of governing equations. However, analytical derivation of the mapping correlations is impractical.

With the rapid advancement of computational capabilities, data-driven approaches have been successfully applied across a wide range of problems, excelling at uncovering hidden correlations within data [50,51]. However, finding universal correlations remain challenging, and the generalization performance of data-driven correlations is often questionable. In light of the above discussion, a data-driven approach is integrated with the concurrent coupling scheme for cryogenic tank simulations to identify the data-driven correlations that map the solutions between CFD and nodal simulations. In this way, mapping correlations between nodal and CFD governing equations are established without analytical derivations. To meet the generalization requirement, the data-driven correlations are generated on the fly for each problem during a co-solving period between the CFD and nodal simulations. Consequently, a data-driven concurrent coupling (DCC) approach has been developed, as illustrated schematically in Figure 3.

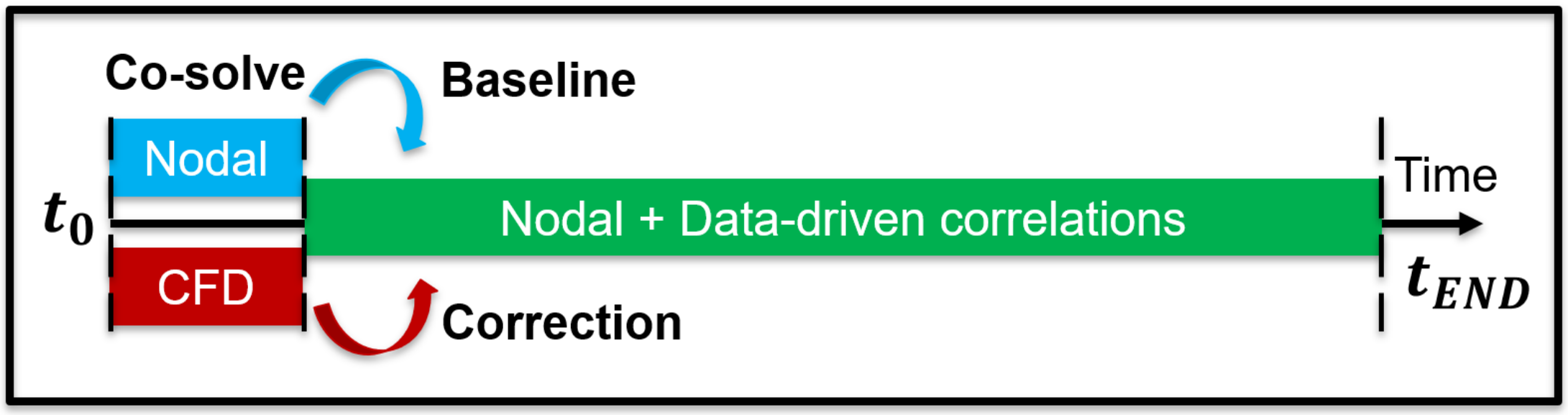

Figure 3. Schematic of the data-driven based coupling approach.

The proposed data-driven based concurrent coupling approach consists of three steps to solve a problem:

1. Co-solving: Model a problem using both nodal and CFD codes independently with identical initial conditions. Then, concurrently run both codes for a small time fraction of the problem.
2. Fitting: The CFD simulation solutions of the short period are volumed averaged following the same resolution defined by the nodal code models. Data-driven correlations are developed to map the nodal code solutions to the volume averaged CFD solutions.
3. Extrapolation: Continue the nodal code simulation until the end of the problem and correct the outcomes using the just-calculated data-driven correlations to generate the final solutions with improved accuracy.

Thus, the DCC approach uses CFD simulation results to enhance the accuracy of the nodal code solutions in an on-the-fly manner.

### 3.1. Physical Significance of Data-Driven Correlations

The existence and physical significance of data-driven correlations, which map the nodal governing equations to the volume averaged CFD governing equations, can be heuristically expounded by analyzing the governing equations of both codes. Taking the energy equation without source term as an example, the CFD solver uses the integrated the N-S equation over each mesh element. The integration of Equation 3 over the $i^{th}$ mesh element with volume $V_{e,i}$ is:

$$\int_{V_{e,i}} \left[\frac{\partial}{\partial t}(\rho u) + \nabla \cdot (\rho u \vec{v})\right] dV = \int_{V_{e,i}} (-\nabla \cdot \vec{q}'' - p\nabla \cdot \vec{v}) dV \,, \tag{10}$$

where the viscous dissipation term is neglected for simplification, and the pressure is defined as a constant within the mesh element. Mutate the time derivative and apply the divergence theorem, the equation becomes:

$$\frac{\partial}{\partial t}\int_{V_{e,i}} \rho u dV + \oiint_{\partial V_{e,i}} \rho \vec{v}\left(u + \frac{p}{\rho}\right) \cdot d\vec{A} + \oiint_{\partial V_{e,i}} \vec{q}'' \cdot d\vec{A} = 0, \tag{11}$$

The first term represents the internal energy of the element and can be defined as $u_{e,i}(t)$:

$$u_{e,i} = \int_{V_{e,i}} \rho u dV \ . \tag{12}$$

The second term represents the energy transfer via mass flow across the element boundaries and can be defined as:

$$\oiint_{\partial V_{e,i}} \rho \vec{v}\left(u + \frac{p}{\rho}\right) \cdot d\vec{A} = -\sum_j fr_j * h_j n_{ij}, \tag{13}$$

where $fr_j(t)$ and $h_j(t)$ are mass flow rate and the specific enthalpy of the fluid across the $j^{th}$ element boundary with the direction indicated by $n_{ij}$.

The last term can be defined similarly with $q_j(t)$ represents the heat flux from the $j^{th}$ element boundary:

$$\oiint_{\partial V_{e,i}} \vec{q}'' \cdot d\vec{A} = -\sum_j q_j n_{ij}. \tag{14}$$

Therefore, the governing equation of the CFD simulation in each mesh element becomes:

$$\frac{\partial}{\partial t} u_{e,i} = \sum_j fr_j * h_j n_{ij} + \sum_j q_j n_{ij}. \tag{15}$$

To compare with the nodal code results, the CFD solutions should be volume averaged according to the corresponding lump region. Therefore, the volume averaged CFD solution for the Lump $k$ with a volume $V_{L,k}$ is:

$$\frac{dU_{CFD,k}}{dt} = \frac{\partial}{\partial t}\left(\frac{1}{V_{L,k}} \sum_i u_{e,i} V_{e,i}\right) = \frac{1}{V_{L,k}} \sum_i \left(\sum_j fr_j * h_j n_{ij}\right) V_{e,i} + \frac{1}{V_{L,k}} \sum_i \left(\sum_j q_j n_{ij}\right) V_{e,i}. \tag{16}$$

Note that $U_{CFD,k}$ is the internal energy for discretized lumps with the same spatial resolution as the corresponding nodal models for comparison. Following the same procedure, the volume averaged CFD solution for a tube can also be obtained. Table 1 shows the comparison of the CFD and nodal solutions for the same control volume.

Table 1. Comparison of the CFD and Nodal Equations for the Same Control Volume

| **Mass Flow Rate for A Tube** | |
|---|---|
| $\frac{d\dot{m}_{CFD,k}}{dt} =$ | $\frac{A_k}{L_k}\left(p_{up} - p_{down} + \frac{m_k \vec{g}}{A_k} - \frac{1}{V_{T,k}} \int_{V_{T,k}} \sum_i^3 \vec{\sigma_i} \frac{A_{e,i}}{A_k}\, dV\right)$ |
| $\frac{d\dot{m}_{Nodal,k}}{dt} =$ | $\frac{A_k}{L_k}\left(p_{up} - p_{down} + C_{1,k} + F_k \dot{m}_k \lvert\dot{m}_k\rvert^{C_{2,k}} + C_{3,k}\dot{m}_k^2 - \frac{K_k \dot{m}_k \lvert\dot{m}_k\rvert}{2\rho_{up} A_k^2}\right)$ |
| **Energy Change for A Lump** | |
| $\frac{dU_{CFD,k}}{dt} =$ | $\frac{1}{V_{L,k}}\sum_i\left(\sum_j fr_j * h_j n_{ij}\right)V_{e,i} + \frac{1}{V_{L,k}}\sum_i\left(\sum_j q_j n_{ij}\right)V_{e,i}$ |
| $\frac{dU_{Nodal,k}}{dt} =$ | $\sum_j \dot{m}_{kj} * h_{k/j} n_{kj} + \sum_j k_{eff,kj}\left(T_{L,j} - T_{L,k}\right)$ |

Upon comparing the above equations, it is apparent that both codes consider the same physical phenomena, but the spatial resolutions of the right-hand side terms are different. In momentum equations, CFD codes use the shear stress tensor, which is affected by both viscosity and flow conditions inside the tube, as shown in Equation 2. However, nodal codes, lacking information about flow distributions inside, rely on simplified models with empirical coefficients to approximate the effects, such as the friction coefficient and the K factor. Similar approximations exist in the energy equations. CFD simulations divide the domain into multiple elements, providing temperature distributions affected by geometry, velocity field, and pressure field, thus accounting for heat transfer in every element. In contrast, nodal codes yield a single value for a lumped region and use a single heat transfer coefficient $k_{eff}$ to approximate the accumulated heat transfer effects. Therefore, the accuracy of nodal code simulations heavily depends on empirical models and coefficients. Selecting these coefficients requires significant expertise, and, at times, suitable empirical correlations may not exist.

To make the solutions of the two codes equivalent (identical), the most straightforward idea is improving the accuracy of the empirical correlations for equivalent parameters, such as the effective heat transfer coefficient and the friction coefficient. Consider the difficulty and limitations of the experiments, CFD solutions can be used to fit new empirical correlations. The domain overlapping coupling approach has been developed based on that idea, which uses the CFD solutions to correct some coefficients, such as the heat transfer coefficient and the friction coefficient of the nodal code used in every time step. Grunloh and Manera [33] successfully applied the domain overlapping approach to several scenarios. They use the CFD solutions to correct the friction coefficients of the nodal code in the overlapped domain and introduced "Velocity Matching Faces (VMFs)" to enforce the consistency of the velocity at the coupling edges. Zhang and Sanchez [35] reviewed previous domain overlapping based coupling work and identified the inherent correlations between different terms of the governing equations. For example, the pressure drops and the velocities can be related by a single equation, thus, correcting the pressure and velocity independently can lead to unsatisfaction of the equation. Therefore, Zhang and Sanchez concluded that negligence of the inherent correlations between variables is the primary defect in the previous domain overlapping work. They developed a new approach to correct multi-coefficients together but that increased the complexity and needed to be carefully designed for specific problems.

In response to these challenges, our proposed coupling approach takes a different route. Instead of addressing individual coefficients, each governing equation of the nodal code is treated as a cohesive unit. As shown in Equation 17, conversion functions $g_k(x)$ are developed to convert the nodal code governing equations into the volume averaged CFD governing equations for the corresponding lumps.

$$\frac{dU_{CFD,k}}{dt} = \frac{1}{V_{L,k}}\sum_i\left(\sum_j fr_j * h_j n_{ij}\right)V_{e,i} + \frac{1}{V_{L,k}}\sum_i\left(\sum_j q_j n_{ij}\right)V_{e,i}$$
$$= \mathcal{g}_k\left(\frac{dU_{Nodal,k}}{dt}\right) = \mathcal{g}_k\left(\sum_j \dot{m}_{kj} * h_{k/j}n_{kj} + \sum_j k_{eff,kj}\left(T_{L,j} - T_{L,k}\right)\right) \quad (17)$$

However, analytically deriving these functions using the above equations is impractical. Therefore, in the proposed DCC approach, data-driven based correlations are used. Equation 18 illustrates the conversion function for the internal energy of Lump k.

$$\frac{dU_{CFD,k}}{dt} = \mathcal{g}_k\left(\frac{dU_{Nodal,k}}{dt}\right) \approx G_k\left(\frac{dU_{Nodal,k}}{dt}\right). \quad (18)$$

The theoretically existing conversion function $\mathcal{g}_k(x)$ is approximated by the data-driven correlation $G_k(x)$ based on the co-solved CFD and nodal code solutions. Equation 18 defines the conversion function at every time step. However, for better stability and representativeness, the data-driven correlations are developed time-independently as the 'averaged' conversion correlation based on the co-solved period data and are used for an extended period, which is detailed in the following sections. Thus, the data-driven correlations are suitable for long-term, slow-changing, and periodic problems, which are typical conditions in cryogenic tank pressure control problems. In contrast, non-periodic, fast-changing problems may limit the extrapolation range of the data-driven correlations. In this way, the inherent relationships between different terms are covered by the data-driven correlations and the conversion functions are effectively represented while eliminating the need for a substantial increase in complexity. Thus, the time evolution of interested variables is directly converted into corresponding volume averaged CFD solutions.

### 3.2. Data-Driven Correlations and the Extrapolation Error

Once the data-driven correlations are obtained, they are employed for an extended period of time. Thus, it is imperative to estimate the extrapolation error to determine the extrapolation limit. Taking the temperature evolution as an example, if the temperature at time step $i$ is solved, for the next time step $i+1$, the temperature is:

$$T_C^{i+1} = T_C^i + Hr_C^i * \Delta t, \quad (19)$$

$$T_N^{i+1} = T_N^i + Hr_N^i * \Delta t, \quad (20)$$

where $T_C$ and $T_N$ represent lump temperatures obtained from CFD and nodal code simulations, respectively. $Hr^i$ is the temperature changing rate at time step $i$ and the time step is $\Delta t$. The DCC approach first co-solve the problem from time step 0 to $n$, and the data-driven correlation $\mathcal{F}\left(T_N^i\right)$ fitted based on the obtained data satisfy:

$$T_C^i \approx \mathcal{F}\left(T_N^i\right), \qquad 0 \le i \le n. \quad (21)$$

At the time step $n$, consider the difference between $T_C^n$ and $\mathcal{F}(T_N^n)$ as $e_{f\mathrm{n}}$, then:

$$T_C^n = \mathcal{F}(T_N^n) + e_{fn} = T_{DCC}^n + e_{fn}. \quad (22)$$

For the time step $n + 1$, the DCC approach result $T_{DCC}^{n+1}$ is:

$$T_{DCC}^{n+1} = \mathcal{F}(T_N^{n+1}). \quad (23)$$

Different orders of polynomial functions are compared as the data-driven correlation function $\mathcal{F}(T_N^i)$, with a detailed derivation of the general polynomial data-driven corrections provided in the Appendix. The results show that higher-order polynomials correlations will introduce $T_N^n \Delta t$ into the correction terms, which grows fast especially when the temperature is high. Consequently, a linear function is selected as the data-driven correlation function $F(T_N^i)$:

$$\mathcal{F}(T_N^i) = a_0 + a_1 T_N^i, \;\; 0 \le i \le n, \quad (24)$$

where $a_0$ and $a_1$ are parameters fitted from the co-solved data. Substitute the linear equation into Equation 23,

$$T_{DCC}^{n+1} = a_0 + a_1 * T_N^{n+1}. \quad (25)$$

Substitute Equation 20 to replace $\mathrm{T_N^{n+1}}$, the DCC approach solution at time step $n + 1$ is:

$$T_{DCC}^{n+1} = T_{DCC}^n + a_1 * Hr_N^n * \Delta t. \quad (26)$$

Apply the data-driven correlation to time step $n + m$, the DCC approach solution $T_{DCC}^{n+m}$ is:

$$T_{DCC}^{n+m} = \mathcal{F}\left(T_N^n + \sum_{i=n}^{i=n+m-1}\left(Hr_N^i * \Delta t\right)\right) = a_1 * T_N^n + a_0 + a_1 * \left[\sum_{i=n}^{i=n+m-1}\left(Hr_N^i * \Delta t\right)\right]. \quad (27)$$

Substitute Equation 24 to Equation 27,

$$T_{DCC}^{n+m} = \mathcal{F}(T_N^n) + a_1 * \left[\sum_{i=n}^{i=n+m-1}\left(Hr_N^i * \Delta t\right)\right] = T_{DCC}^n + a_1 * \left[\sum_{i=n}^{i=n+m-1}\left(Hr_N^i * \Delta t\right)\right]. \quad (28)$$

The CFD solution at time step $n + m$ is:

$$T_C^{n+m} = T_C^n + \left[\sum_{i=n}^{i=n+m-1}\left(Hr_C^i * \Delta t\right)\right], \quad (29)$$

Therefore, the error of the DCC approach at time step $n + m$ is:

$$e_{DCC}^{n+m} = \sum_{i=n}^{i=n+m-1}\left(a_1 * Hr_N^i - Hr_C^i\right)\Delta t - e_{fn}, \quad (30)$$

The fitting error $e_{f\mathrm{n}}$ comes from the fitting process of the first co-solved n-point data. It is a fixed constant and is generally negligible ($e_{f\mathrm{n}} \approx 0$) if the fitting algorithm is carefully selected. Therefore, this term is neglected in the following discussion of the extrapolation error. Since both $Hr_N^i$ and $Hr_C^i$ are continuously changing, according to the Lagrange mean value theorem, there exist mean values for the temperature changing rate, $\overline{Hr_N}$, and $\overline{Hr_C}$, within the time step from $n$ to $n+m$, which satisfy that:

$$\overline{Hr_N} = \frac{\left[\sum_{i=n}^{i=n+m-1}(Hr_N^i * \Delta t)\right]}{m * \Delta t}, \tag{29}$$

$$\overline{Hr_C} = \frac{\left[\sum_{i=n}^{i=n+m-1}(Hr_C^i * \Delta t)\right]}{m * \Delta t}. \tag{30}$$

Thus, the extrapolation error of the DCC approach solution is:

$$e_{DCC}^{n+m} = (a_1\overline{Hr_N} - \overline{Hr_C})m\Delta t, \tag{31}$$

Equations 27 and 33 indicate the underlying physical explanation of the linear data-driven correlations. Specifically, the slope $\mathrm{a}_1$ is trying to compensate for the difference of the average heating rate between the two codes, whereas the intercept $\mathrm{a}_0$ serves to preserve the continuity at time step n between the DCC solutions and the CFD solutions.

From Equation 33, the extrapolation error of the DCC approach comes from the difference between the average temperature changing rate of the two solutions. If the difference of the two average temperature changing rate can be estimated, the extrapolation error will exhibit a linear increase over time (m).

In practice, the estimation of the $(a_1\overline{Hr_N} - \overline{Hr_C})$ term is addressed together with the selection of the duration for the co-solved period. This process involves with four steps:

1. First, an initial co-solve period of $n$ time steps is selected and proceeded. The co-solved data is divided into two datasets: a fitting dataset for the first $m_f$ time steps and a validation dataset for the rest of $m_V$ time steps.
2. The data-driven correlations are initially fitted using the fitting dataset. These correlations are then used with nodal codes solutions to predict the validation dataset. The difference between the predictions and the actual values at the end of the $m_V$ time step in the validation dataset, denoted as $e_{\mathrm{V}}^{m_f+m_V}$, can be calculated exactly using Equation 34, and meanwhile is correlated to Equation 33:

$$e_V^{m_f+m_V} = (T_{DCC}^{m_f+m_V} - T_C^{m_f+m_V}) \approx (a_1\overline{Hr_N} - \overline{Hr_C})m_V\Delta t. \tag{324}$$

   Thus, the term $(a_1\overline{Hr_N} - \overline{Hr_C})$ in Equation 33 can be estimated by the validation dataset during co-solve period, and then be used to estimate errors after the co-solve period.

3. If $e_{\mathrm{V}}^{m_f+m_{\mathrm{V}}}$ is substantially large, which means the co-solved period is not long enough to capture necessary information, we need to go back to Step 1 to extend $n$ for extra co-solved data and following the steps again. If $e_{\mathrm{V}}^{m_f+m_{\mathrm{V}}}$ is small enough, the error between the DCC predictions and the actual values for the following $m$ time steps of the problem can be assumed linearly increase with time using the proportionality factor $(a_1\overline{Hr_N} - \overline{Hr_C})$ just determined by the validation dataset. Therefore, the long-term extrapolation error $e_{DC}^{n+m}$ can be estimated as:

$$e_{DCC}^{n+m} = (a_1\overline{Hr_N} - \overline{Hr_C})m\Delta t = \frac{m}{m_V}e_V^{m_f+m_V}. \tag{335}$$

4. After that, the validation dataset will be incorporated into the fitting dataset, facilitating the refinement of the final data-driven correlations. Due to the inclusion of additional data, the final data-driven correlations are improved, and one can expect the extrapolation error in the final solution to be smaller than the initially estimated error.

Thus, the extrapolation error can be estimated in a conservative manner. The estimation of the extrapolation error for other interested variables, such as the pressure, can be obtained following the same steps.

## 4. Results and Discussions

In this section, the performance of the DCC approach is demonstrated in improving the accuracy of nodal simulations using short period co-solving CFD simulations. The DCC approach is first evaluated using synthetic tests and subsequently validated against experimental data.

In the synthetic tests, CFD simulations are assumed to be accurate and are treated as benchmarks. A short period of co-solving CFD data is used to correct the nodal simulations, and the resulting DCC solutions are compared against the complete CFD solutions for the entire problem. For experiment validations, a short period of CFD simulation data from the literature is used to correct the nodal simulations, and the DCC results are compared to corresponding experimental measurements.

### 4.1. Synthetic Test

The synthetic tests include two representative scenarios in a two-phase cryogenic tank, namely, thermal stratification and periodic mixing. In CFD simulations, the Laminar model is applied to the first problem, while the SST k-omega model is selected for the second one. The VOF approach is used to track the interface of the two-phase fluid.

Although detailed validation of the CFD model is beyond the scope of this study, similar configurations have been validated by Kassemi et al. [52], who compared CFD predictions to experimental results across various cryogenic tank geometries. It should be noted that although Kassemi et al. [52] used a different interface modeling approach than the Lee model applied here, both models contain tunable parameters that enable alignment of the CFD simulations with experimental measurements. For the purposes of this study, the CFD solutions are assumed to be sufficiently accurate.

#### 4.1.1. Thermal Stratification during Tank Self-Pressurization

In this section, the DCC approach is implemented for a self-pressurization problem with external heat fluctuation. A cryogenic tank model is shown in Figure 4 (a), which is cylindrical with a 1-meter height and inner diameter. The tank is initially filled with liquid and vapor nitrogen with a 75% liquid fill level, saturated at 100 kPa (77.24 K). The bottom and side walls are adiabatic, with the heat flux only being applied to the tank's lid. Within 500 seconds of fluid time, the heat flux is constant at 50 $W/m^2$ for the first 400 seconds resulting in a heat load of 39.27 W. Then the heat flux starts to fluctuate and becomes stable again until 460 seconds. Figure 5 (b) illustrates the heat load on the tank's lid during the simulation.

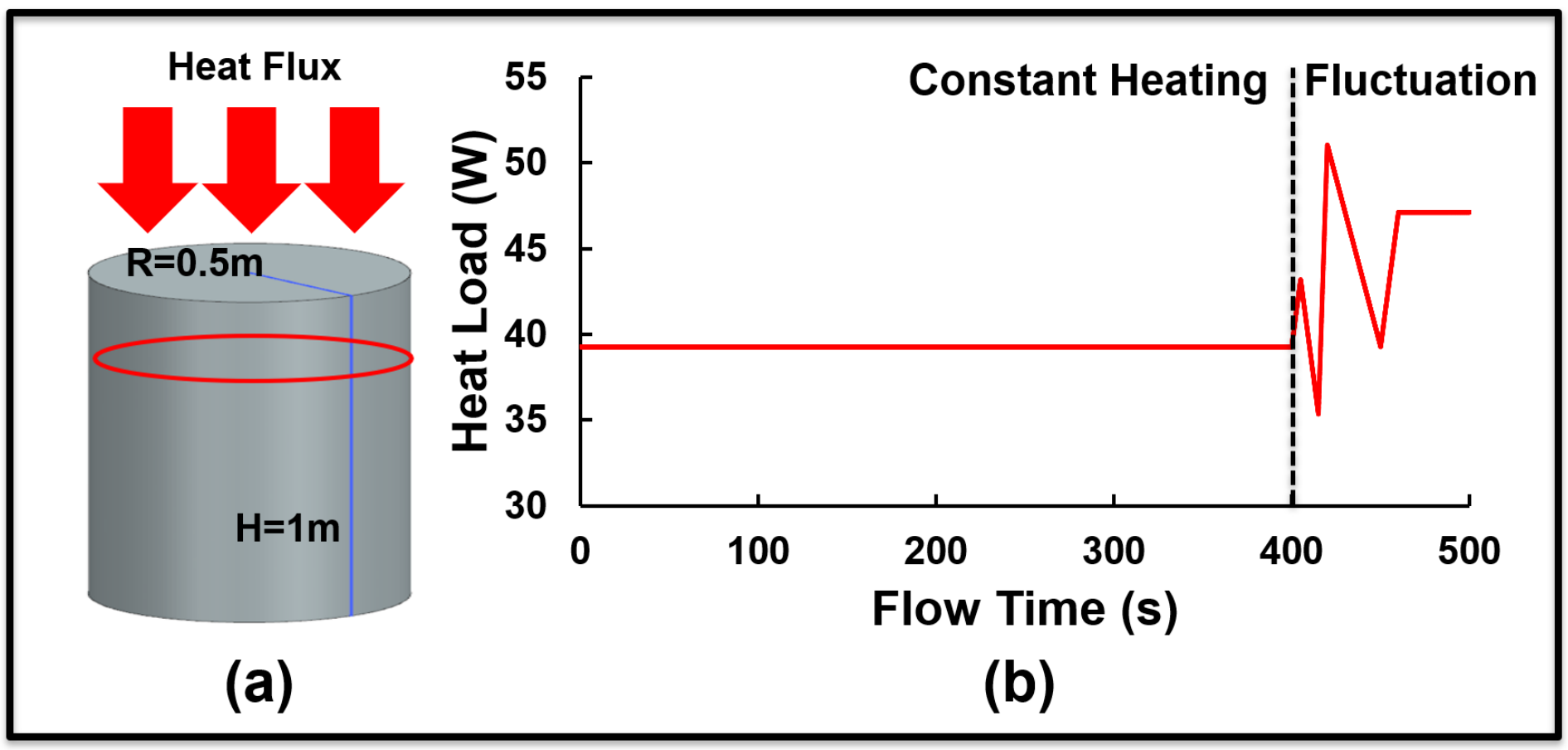

Figure 4. (a) Geometry of the cryogenic tank for the self-pressurization problem. (b) The heat load applied to the tank's lid. The heat load varies after 400 seconds of fluid time.

Figure 5 shows the CFD solutions of temperature contours and the stream functions at various time points. The thermal stratification within the system emerges due to the external heat source to the tank's lid, initially manifesting at the upper region and gradually descending. Thus, the hottest part of the vapor is at the top, close to the wall, where external heat enters the tank, while the coldest vapor is at the bottom interface, where heat is lost to the liquid and condensation ensues. Notably, the vapor region exhibits the most prominent thermal stratification, while the liquid region maintains a nearly uniform temperature profile due to the natural convection mixing and the high heat capacity. The stream functions give the velocity directions in different locations, revealing intricate characteristics arising from both the natural convection and phase changes occurring at the interface. In Figure 5(a), during the initial stages, the fluid's movement is primarily driven by phase changes near the interface, therefore, the vortex induced by the phase change dominates in the vapor region. As the natural convection currents gain strength upon external heat, they compress the two vortices of the phase change significantly against the interface. Since the liquid does not directly contact the heat source, heat transfer is primarily governed by conduction through the interface. Consequently, two symmetrical vortices gradually form, occupying most of the liquid region, indicating the relatively weaker influence of phase changes compared to conduction. At the end of the simulation, depicted in Figure 5(f), several small vortices and pronounced thermal stratification exist in the vapor region. In contrast, the liquid region maintains a near-uniform temperature with two dominated vortices driven by the buoyancy force. These observations underscore the 3D nature of cryogenic tank behaviors, which holds significant relevance for the pressurization process and presents considerable challenges for accurate simulation in nodal codes.

It is important to note that uniform initialization for the CFD simulations of cryogenic tanks is a common compromise due to the insufficient information available from experimental data, such as velocity and temperature distributions [23]. However, using a uniform initial condition is not suitable for CFD simulations because it lacks both the thermal and velocity boundary layers. As a result, an unstable period in the fluid field within the cryogenic tank is observed at the beginning of the CFD simulation, as depicted in Figure 5(a). This initial state exhibits distinct flow patterns when compared to the others in Figure 5. Stewart and Moder [8] attempted to address this issue by simulating the preparation process of the cryogenic tank experiments to obtain more realistic initial conditions which, however, is very time consuming. Therefore, in this study, the first 100 seconds of data are excluded for the following analysis to allow the fluid field of the CFD simulation to reach a stable developing status, enhancing the reliability of our solutions.

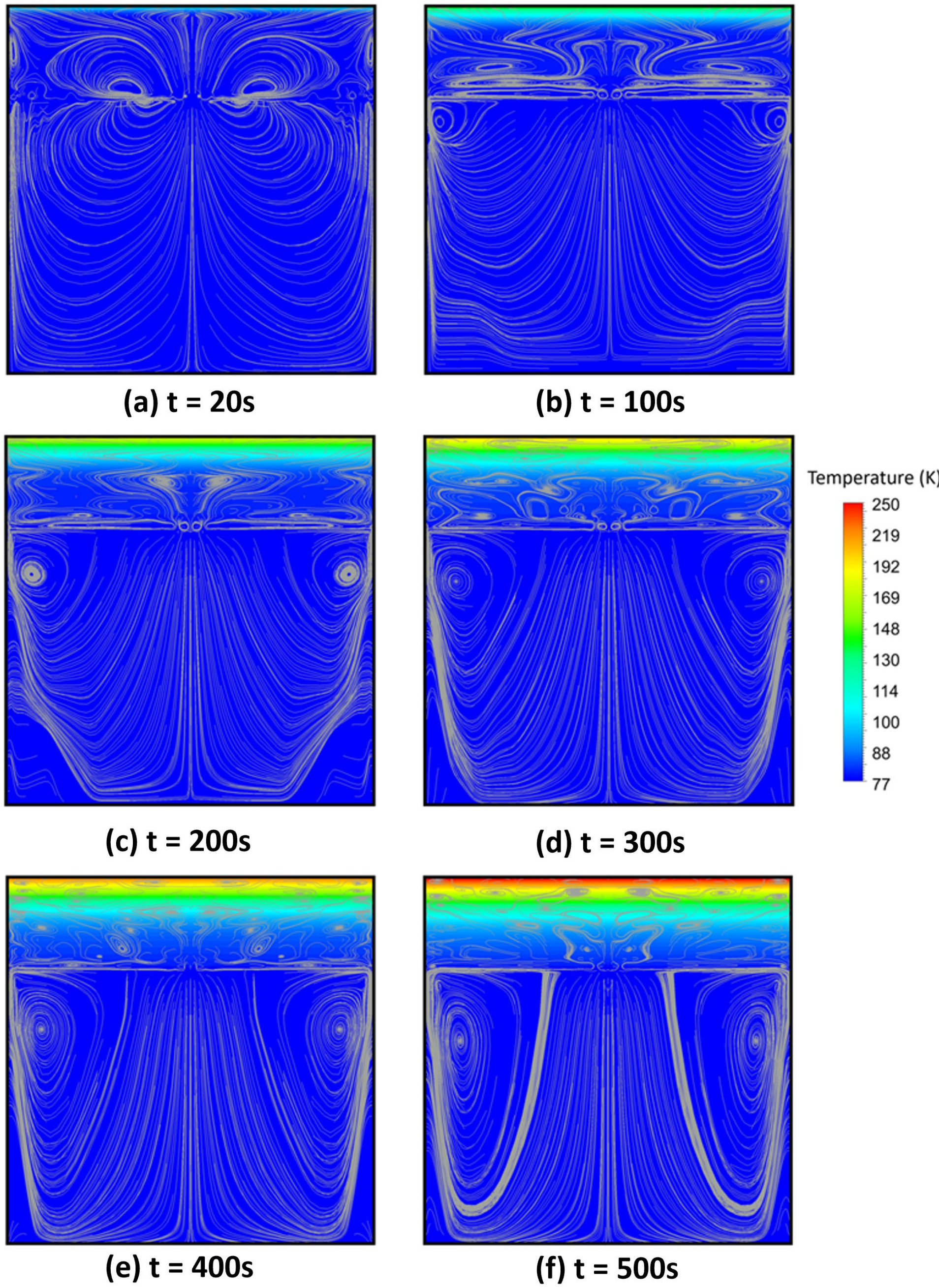

Figure 5. CFD solutions of the self-pressurization process. The background is the temperature contour plot, and the gray lines are stream functions.

To compare CFD solutions with nodal code solutions, the computational domain of the CFD model is divided vertically into nine regions based on the nodal code model's geometry. As shown in Figure 6, the volume averaged temperatures of CFD and nodal simulations for the top three regions are compared, where the thermal stratification is the most pronounced. The temperatures of the three regions increase continuously while the discrepancy between the two solutions is also growing, especially in Region 1, where the thermal stratification is the most significant and the simulation solutions of the nodal code is the most inaccurate. Note that after 400 seconds of fluid time, the temperature fluctuated due to the changing heat flux boundary conditions shown in Figure 4(b).

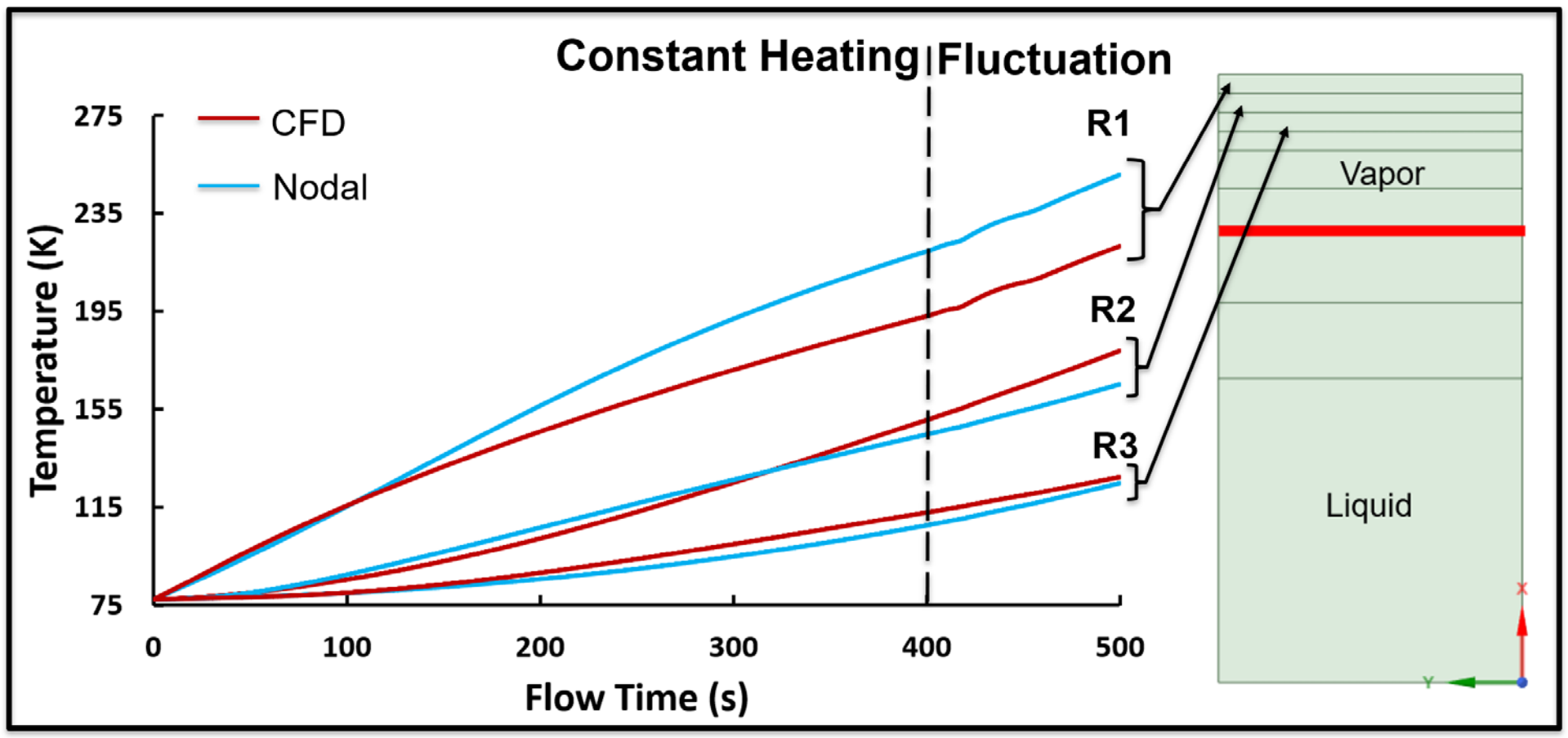

Figure 6. CFD and nodal code solutions of the volume averaged temperature of the top three regions.

As discussed in Section 3, the DCC approach uses the co-solve data to develop case-specific correlations on the fly for interested variables. In this case, the correlations are developed to convert the nodal code solutions of the lump temperature into the corresponding volume-averaged CFD solutions, thereby compensating for the missing 3D information and improving accuracy. Since the first 100 seconds of data are discarded, the data from 100 to 150 seconds are used as the co-solved data. Thus, for the DCC approach, the CFD simulation only runs up to 150 seconds instead of the entire 500 seconds. The 50 seconds co-solved data is split into a 40-second fitting dataset and a 10-second validation dataset. Figure 7 shows the comparison and the fitting process for the top three regions. In Figure 7(e) to (f), the horizontal axis is the nodal code solutions, and the vertical axis is the CFD solutions. The phase curves are plotted by pairing nodal and CFD solutions $(N(t), CFD(t))$ at each time step. Then, linear regression is used to obtain the approximation functions $\mathcal{F}$ discussed in Equation 18, which will map the map the nodal simulation solutions $N(t)$, to match the volume averaged CFD solutions $CFD(t)$. The obtained data-driven correlations are lump (spatial) dependent but time independent. This fitting process is applied to all regions and the data-driven correlations are obtained for each region. Comparing to the top three regions, the difference between CFD and nodal codes solutions is insignificant in the remaining regions, especially for the liquid regions, due to the stronger mixing inside the liquid. To estimate the extrapolation error of the data-driven correlations, take the Region 1 as an example, the data-driven correlation of Region 1 is first applied to the validation dataset and results in a difference of 0.286 K for the 10 seconds of extrapolation. Therefore, assume the same performance for the rest of the 350 seconds extrapolation, it will result in an estimated extrapolation error of 10.01 K. The 10 seconds of validation dataset is then added back to improve the data-driven correlations.

Figure 8 illustrates a comparison of solutions among the DCC approach, CFD, and nodal codes for the volume averaged temperature in the top three regions. The red and blue lines depict the solutions from CFD and nodal codes, respectively, while the green cross markers represent the DCC approach solutions. CFD and nodal codes co-solve the first 150 seconds and the data between 100 to 150 seconds are used to find the data-driven correlations. The DCC approach extrapolates correlations to refine nodal code solutions up to 500 seconds, resulting in improved accuracy compared to nodal code results, particularly in the Region 1, characterized by the highest temperature increase. The DCC results after 400 seconds demonstrate the robustness of the data-driven correlations to the fluctuations of the boundary conditions. That comes from the similar structure between the governing equations of the CFD and nodal codes, resulting in a similar response to changes in boundary conditions and a relative consistent difference between the two solutions. Therefore, the evolution curves of the CFD and nodal codes solutions could exhibit a consistent trend despite fluctuations in boundary conditions, and the data-driven correlations are expected to remain effective over an extended period. The difference between the DCC approach results and the volume averaged CFD solutions in the Region 1 at 500 seconds is 1.04 K, which is smaller than the estimated extrapolation error, demonstrating the effectiveness of the proposed conservative extrapolation error estimation approach. The most pronounced improvement occurs in the top Region 1 since the most heat enters that region, making the fluid field in the Region 1 develop fast, resulting in long-term-like behavior. In contrast, the following two regions are still developing due to the relatively short simulation time, and nonlinearity plays a more important role, which shows the limitation of the linearly fitted correlations. However, despite linear correlations being the most fundamental and straightforward to apply, the DCC approach is naturally suitable to accommodate any data-driven approach to finding the mapping correlations from the co-solved data to correct the nodal code solutions. This flexibility positions the DCC approach as adaptable to more advanced fitting algorithms, including the fast-developed neuron networks.

The CFD and nodal codes use 10 hours and 12.6 mins, respectively, to finish the 500 s simulation. The DCC solves the problem in around 3 hours, reducing 70% computational time compared to the CFD simulation.

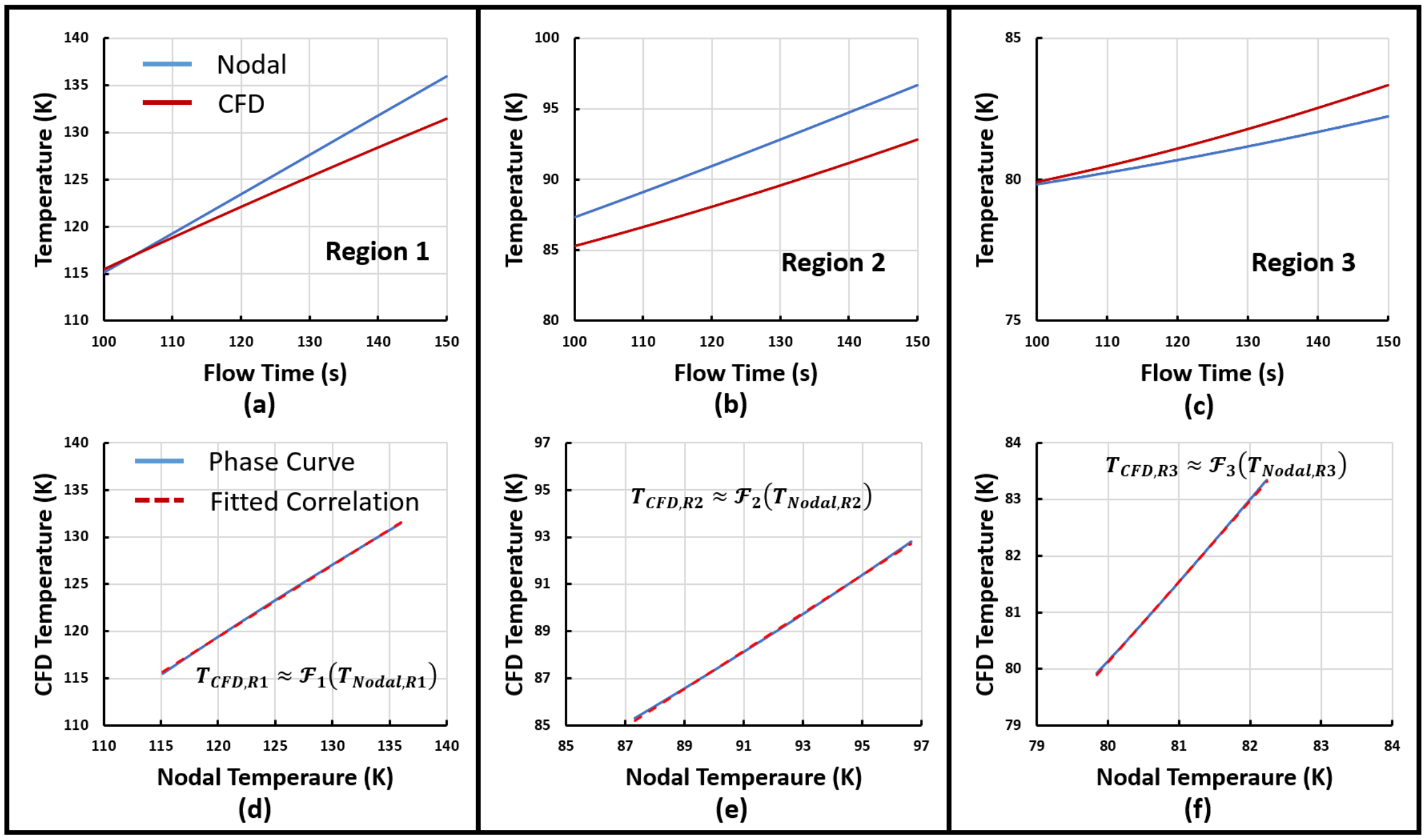

Figure 7. (a)-(c) Co-solved data showing the volume averaged temperature in the top three regions between 100 to 150 seconds. (d)-(f) Phase curves and the fitted correlations ($\mathcal{F}_1$, $\mathcal{F}_2$, and $\mathcal{F}_3$) for the volume-averaged temperature in the top three regions. The phase curves are plotted by pairing nodal and CFD solutions at each time step, resulting in time-independent data-driven correlations.

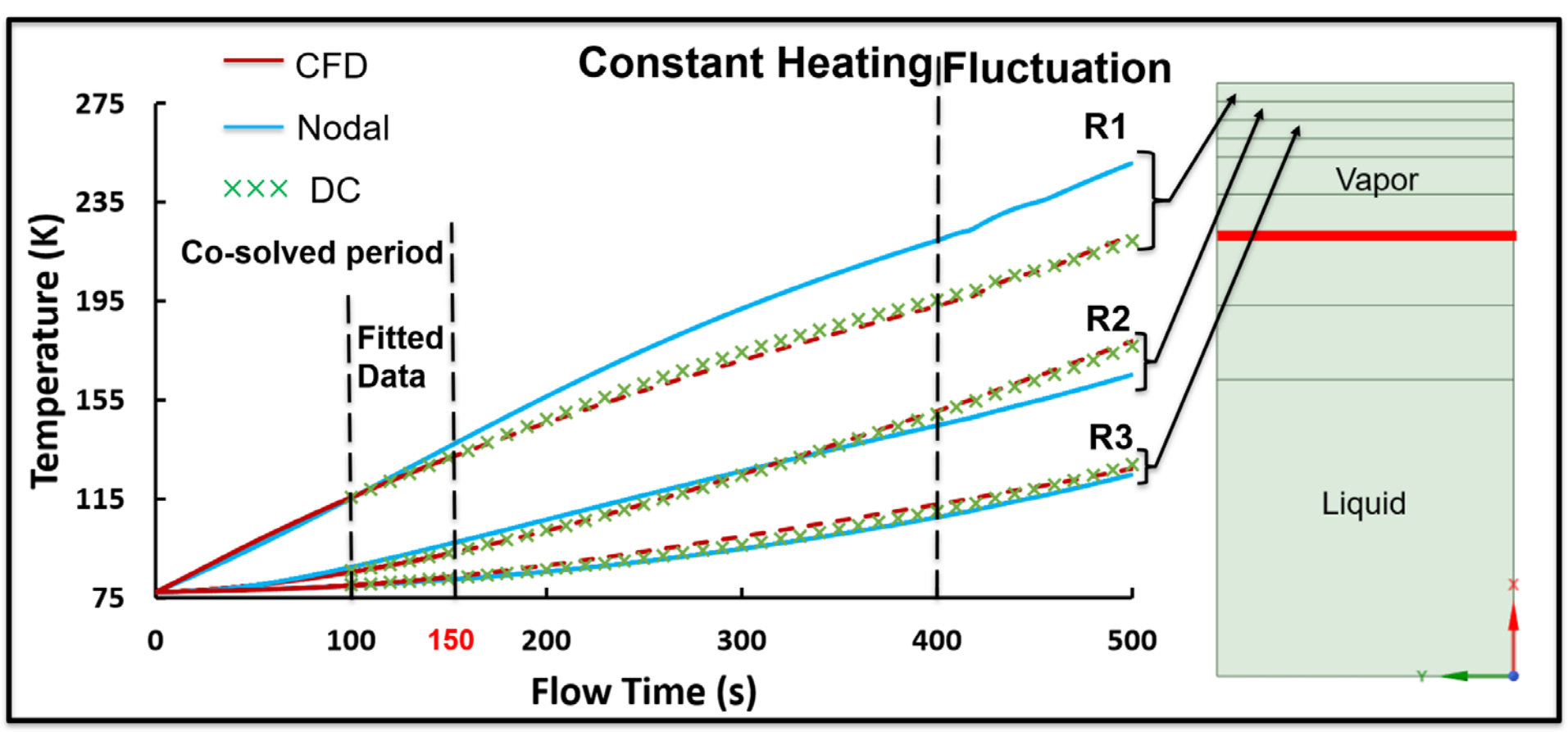

Figure 8. Comparison of volume averaged temperature solutions of different approaches.

It is also worth mentioning that the heat flux used in this demonstration is significantly higher than what would be encountered under realistic conditions, aiming to increase the temperature and pressure changes in a relatively short simulation period. In actual NASA space missions, cryogenic propellant tanks are typically well insulated, and the temperature usually increases only a few Kelvins across several hours of the self-pressurization process [20,23,24]. Consequently, the flow conditions within the cryogenic tanks during self-pressurization change more slowly than those presented in the demonstration problem, resulting in a more linear temperature/pressure evolution curve. This long-term but slow-changing nature of temperature and pressure variations aligns well with the proposed DCC approach. Therefore, it is reasonable to anticipate that the DCC approach would perform even better in such practical applications.

### 4.1.2. Mixing in Heating-depressurization Cycles

In this section, a periodically operated cryogenic tank is used to demonstrate the capability of the DCC approach. The geometry of the periodically operated tank is shown in Figure 9(a) with the identical dimensions of the tank to the one in Section 4.1. The heat flux is fixed at 100 $W/m^2$. The tank is vented with a constant mass flow rate from the valve on the lid of the tank for 1 second every 20 seconds, starting at the fluid time of 100 seconds. Cold liquid with the same mass flow rate flows into the tank through the bottom to conserve the mass of the tank. As shown in Figure 9(b), the grey line represents the mass flow rate, which remains zero when the valves are closed and becomes a constant value during the short venting periods. The blue and red lines represent the tank pressure simulation results from the nodal and CFD codes, respectively. Starting at 100 seconds, the tank pressure sharply decreases for 1 second due to the opening of the venting valve and then continues to increase after the valve is closed. Consequently, the tank pressure evolution curves exhibit similar but not identical cycles.

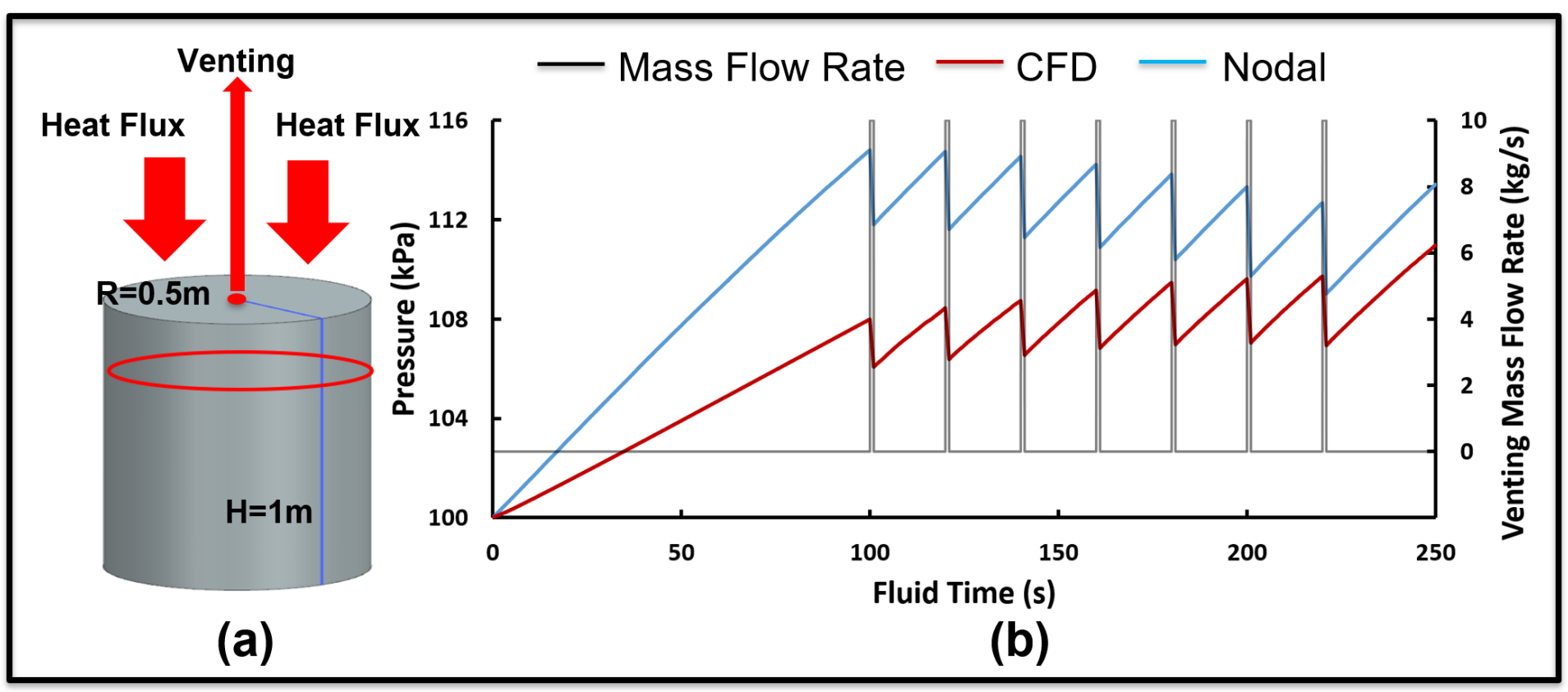

Figure 9. The mass flow rate boundary conditions during the simulation and the corresponding pressure responds from CFD and nodal codes solutions.

Figure 10 shows the CFD solutions of temperature contours and the stream functions near a venting event. In Figure 10(a), the tank undergoes self-pressurization for 20 seconds following the last venting and is ready for the next venting occurrence at t = 160s. Therefore, the vortex arising from the phase change can be observed near the interface, while forced convection within the liquid region is weakening. However, following the venting and subsequent jet injection at t = 170s, as illustrated in Figure 10(b), two new vortices emerge in the vapor region, signifying the dominance of forced convection resulting from the venting process in the vapor region. Simultaneously, forced convection driven by the jet injection encompasses the entire liquid region.

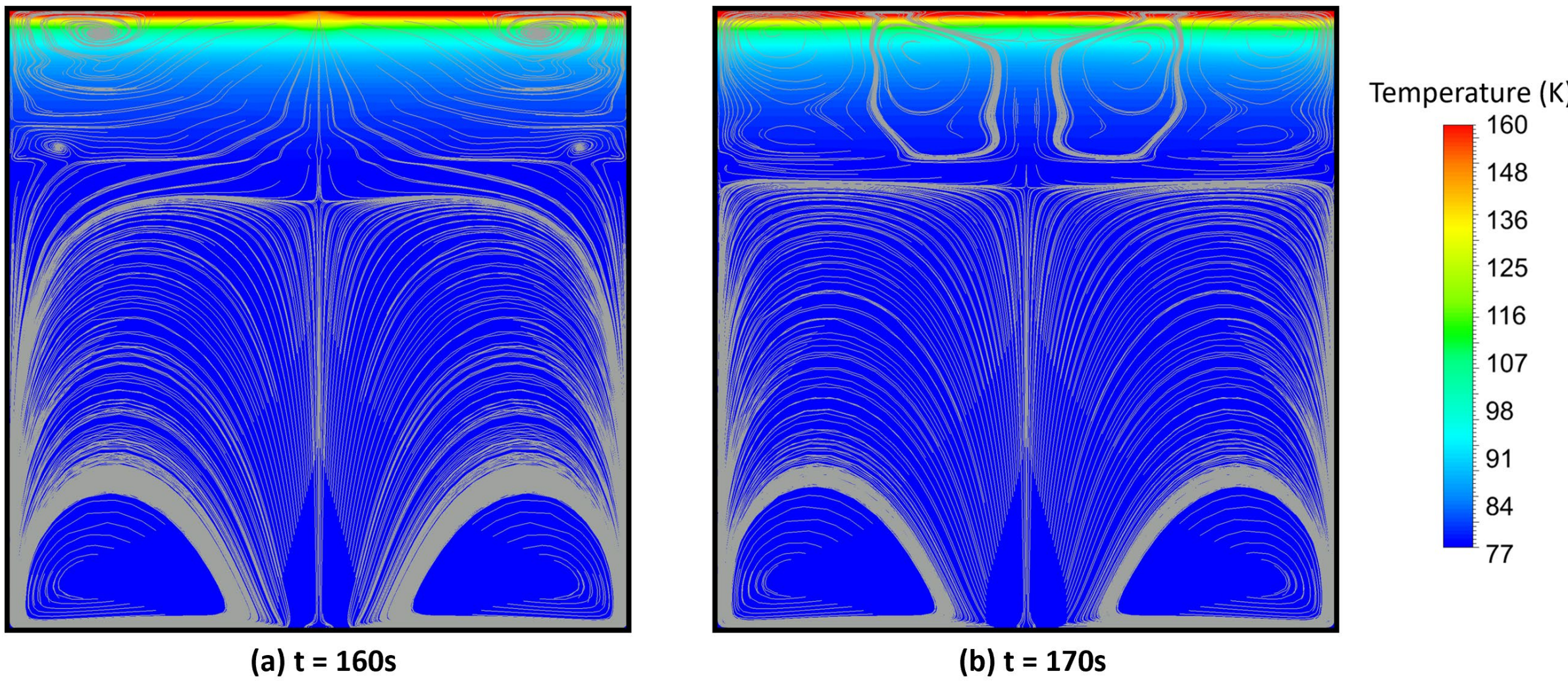

Figure 10.  CFD solutions of the heating-depressurization process. The background is the temperature contour plot, and the gray lines are stream functions.

Figure 11 presents the temperature evolution curves for the top two regions with the same spatial discretization as in Section 4.1. After 100 seconds of pressurization, the temperature curves exhibit periodic behavior due to the venting process. Since the initial 100 seconds of self-pressurization serve to prepare the tank for venting and show no significant differences compared to Problem 1, the focus of this demonstration is on the tank's behavior following the self-pressurization. Thus, the DCC approach uses the CFD and nodal codes to co-solve the problem until the end of the first cycle (120 s). Based on the co-solved data from 100 to 120 seconds, data-driven approach is used to obtain two correlations for venting and pressurization, respectively, in every region, as shown in Figure 12. Due to the fluctuated starting points for different cycles, the data-driven correlations are adapted to maintain the continuity of the curve as discussed in Section 3.2.

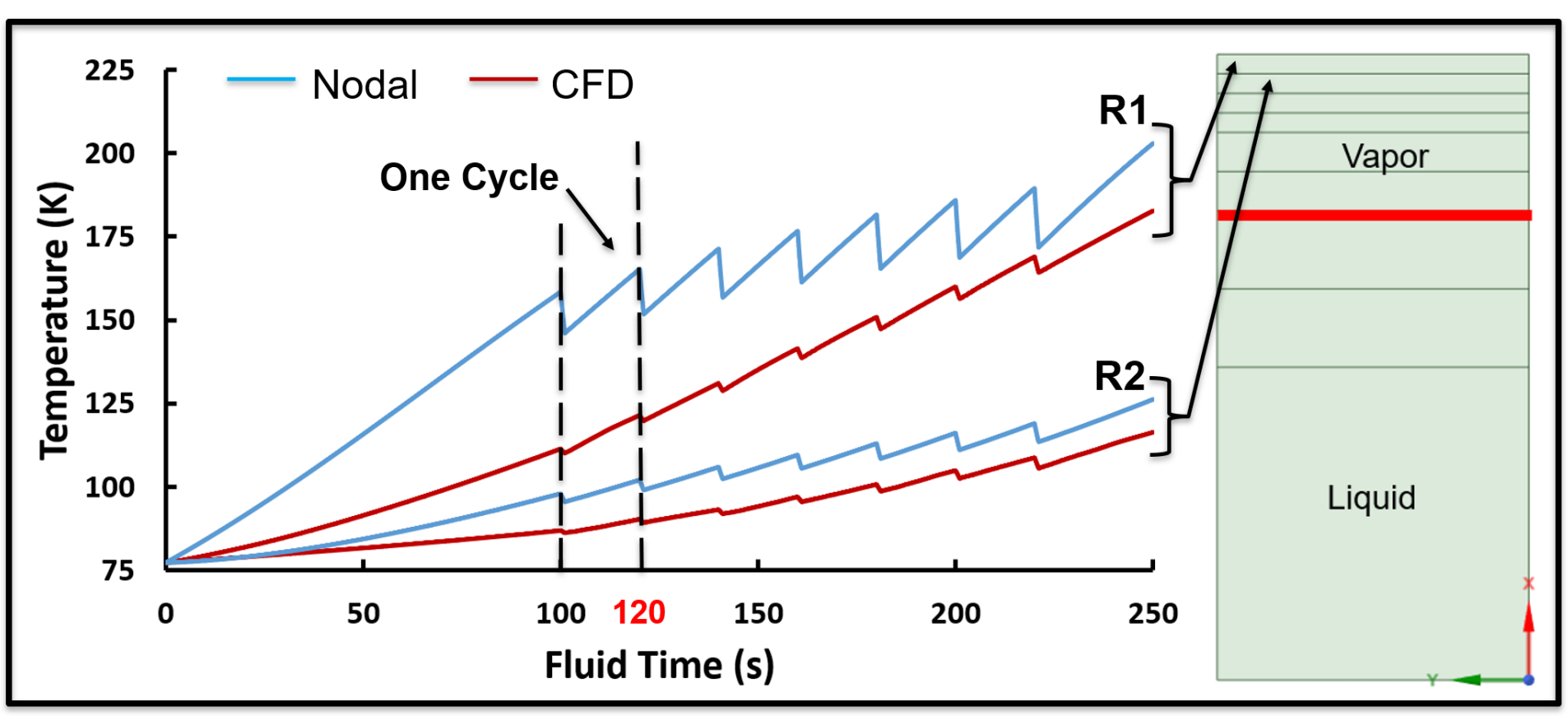

Figure 11. Comparison of the tank volume averaged temperature results between the nodal and CFD solutions.

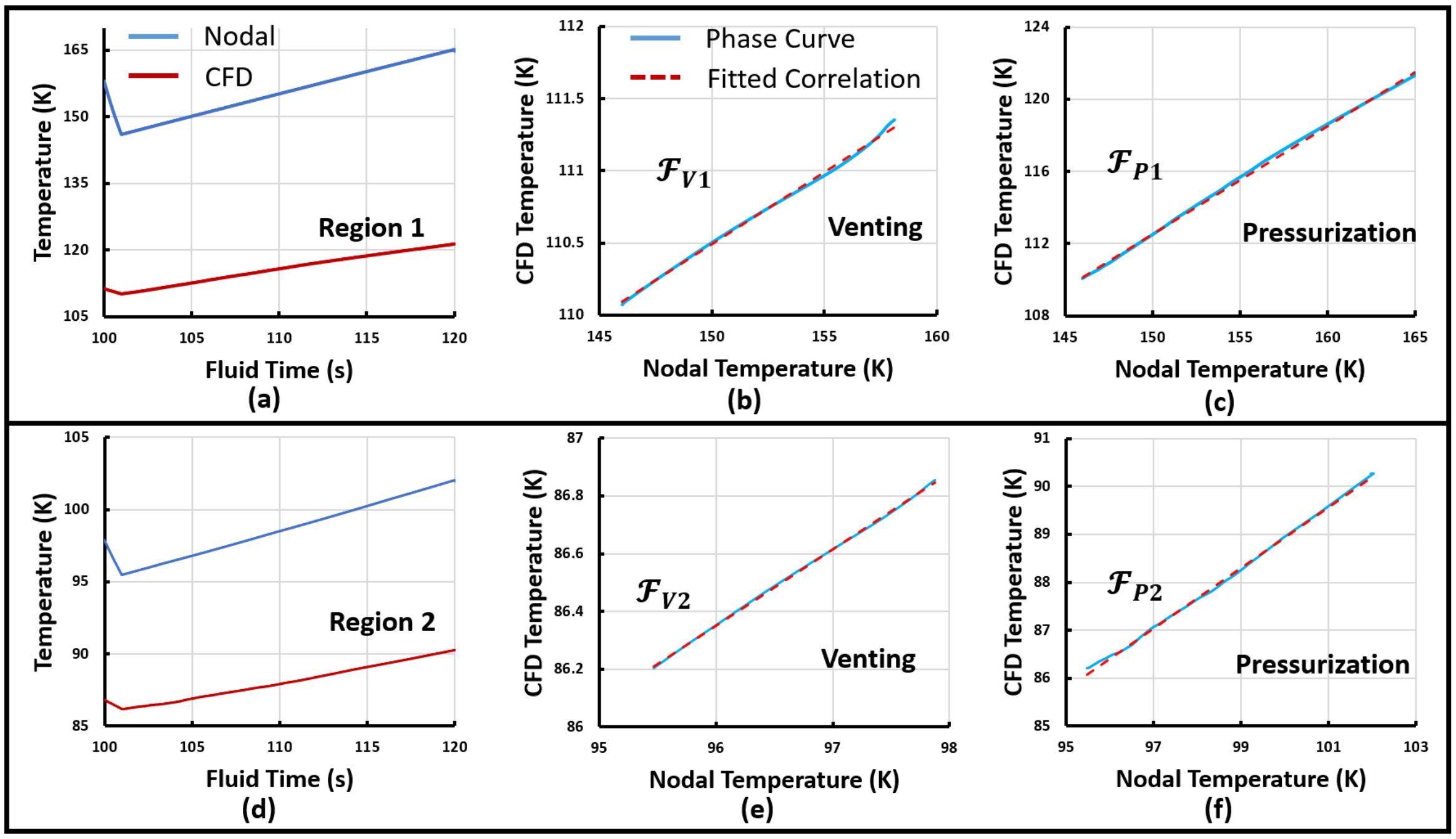

Figure 12. (a) and (d) Co-solved data of the volume averaged temperature in the top two regions between 100 to 120 seconds. (b) and (e) Phase curves and the fitted correlations ($\mathcal{F}_{V1}$, $\mathcal{F}_{V2}$) of the venting process (100-101s) in the two regions. (c) and (f) Phase curves and fitted correlations ($\mathcal{F}_{P1}$, $\mathcal{F}_{P2}$) of the pressurization process (101-120s) in the two regions. The phase curves are plotted by pairing nodal and CFD solutions at each time step, resulting in time-independent data-driven correlations.

Figure 13 compares the volume averaged temperature solutions of the DCC, CFD and nodal codes. The DCC approach obtains the corresponding mapping correlations and extrapolate these until the end of the problem. The green cross markers in Figure 13 represent the DCC solutions which agree with the volume averaged CFD solutions very well. Although the venting-pressurization cycles are not identical, the data-driven correlations obtained using the one cycle co-solved data successfully improve the accuracy of the nodal solutions for six more cycles. Figure 14 presents the relative error of the nodal and DCC solutions to the corresponding volume averaged CFD solutions. In both regions, the DCC approach lower the relative error significantly.

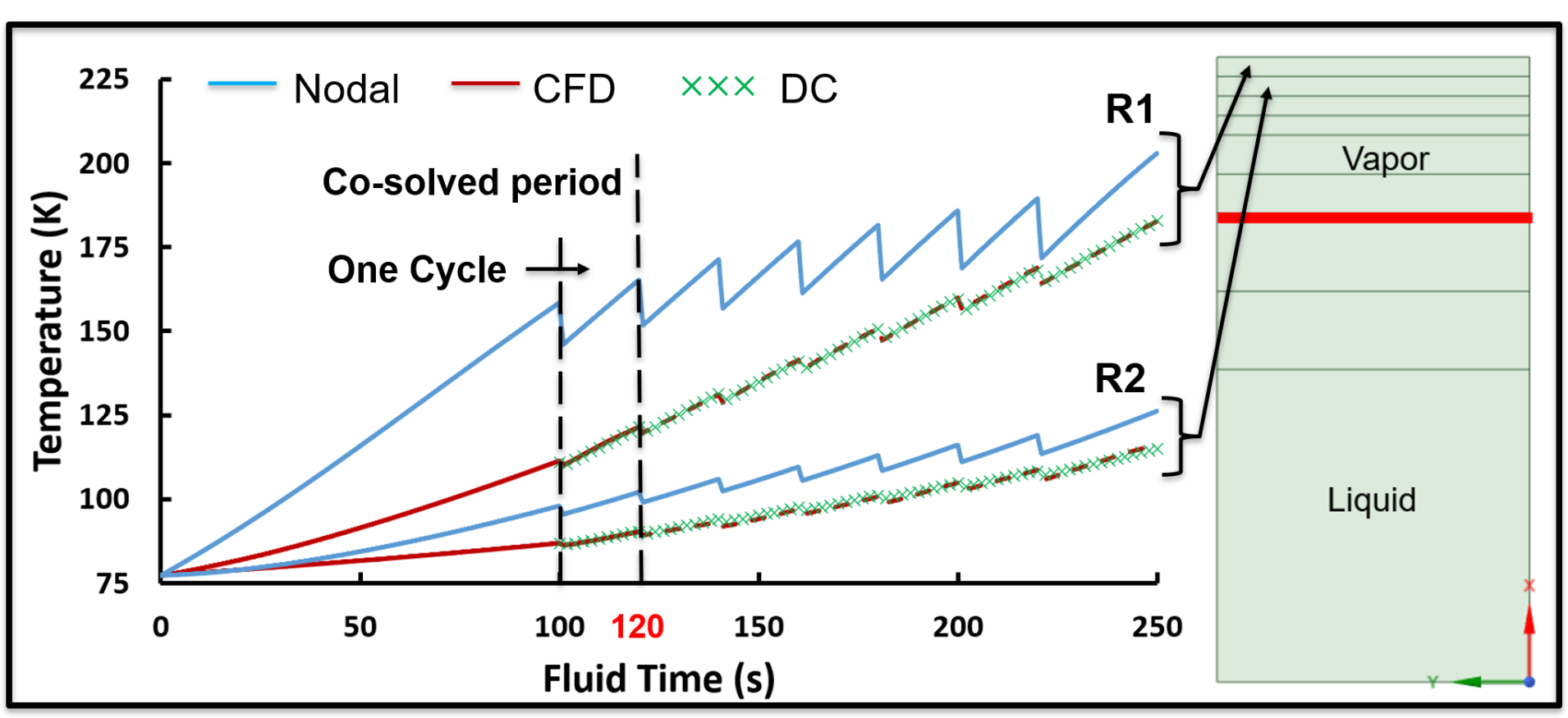

Figure 13. Comparison of the volume averaged temperature solutions between the DCC, CFD and nodal codes of the top two regions of the tank.

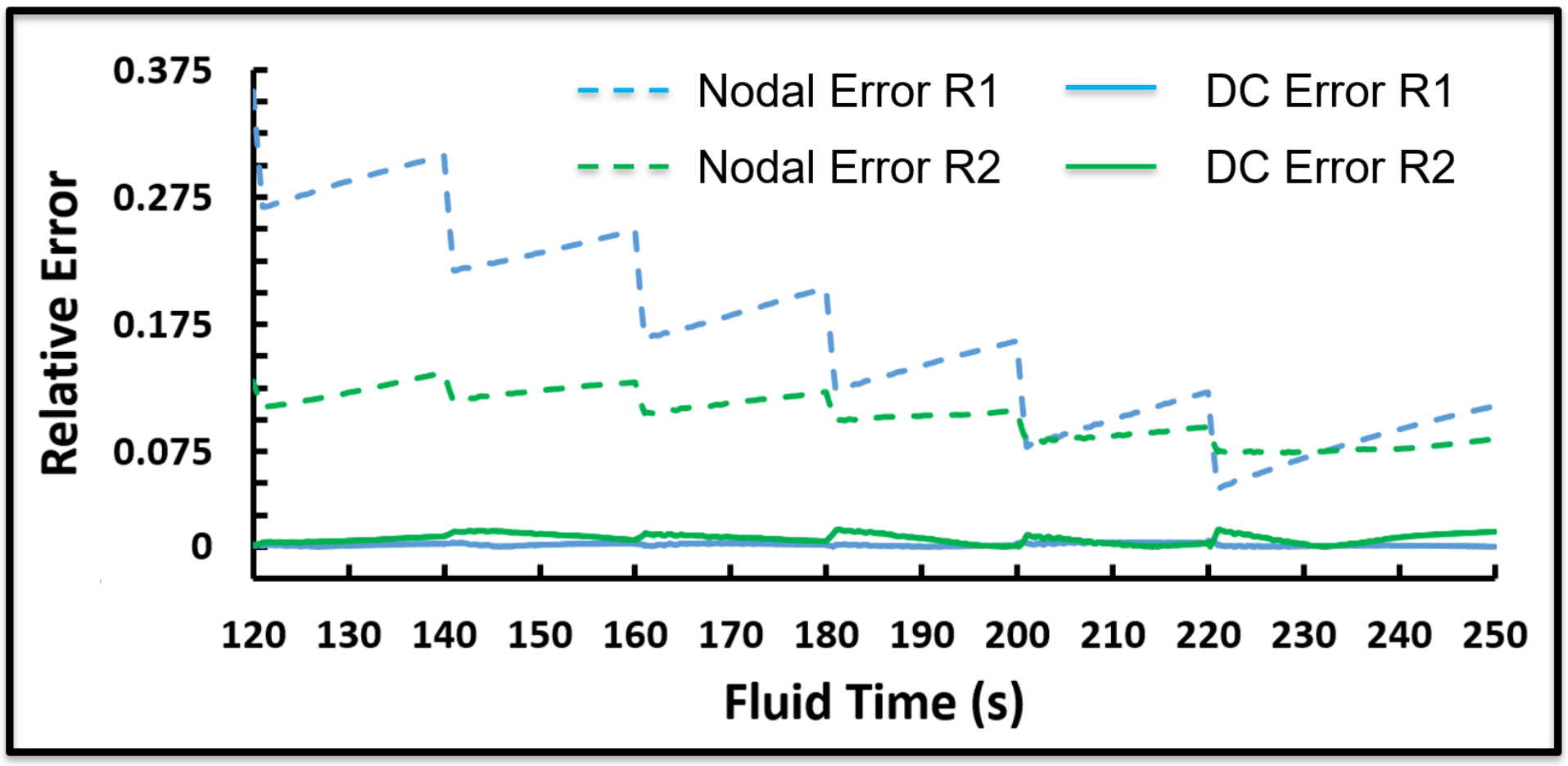

Figure 14. Comparison of the relative errors between the DCC and nodal solutions to the volume averaged CFD solutions of the top two regions of the tank.

Following the same process with the co-solved data, Figure 15 compares the tank pressure solutions obtained using the DCC approach with those from the CFD and nodal solutions. The relative errors of the DCC and nodal solutions compared to the CFD results are also shown with dashed lines. Similar to the volume averaged temperature, although the pressure cycles are not identical, the data-driven correlations maintain good accuracy when extrapolated to six additional cycles. This highlights the key features of the DCC approach: (1) the nodal code can capture the same trends in evolution curves as the CFD code, and (2) the intrinsic differences are compensated by the data-driven correlations. For the periodic mixing problem, the DCC approach reduces computational costs by 52% compared to CFD simulations.

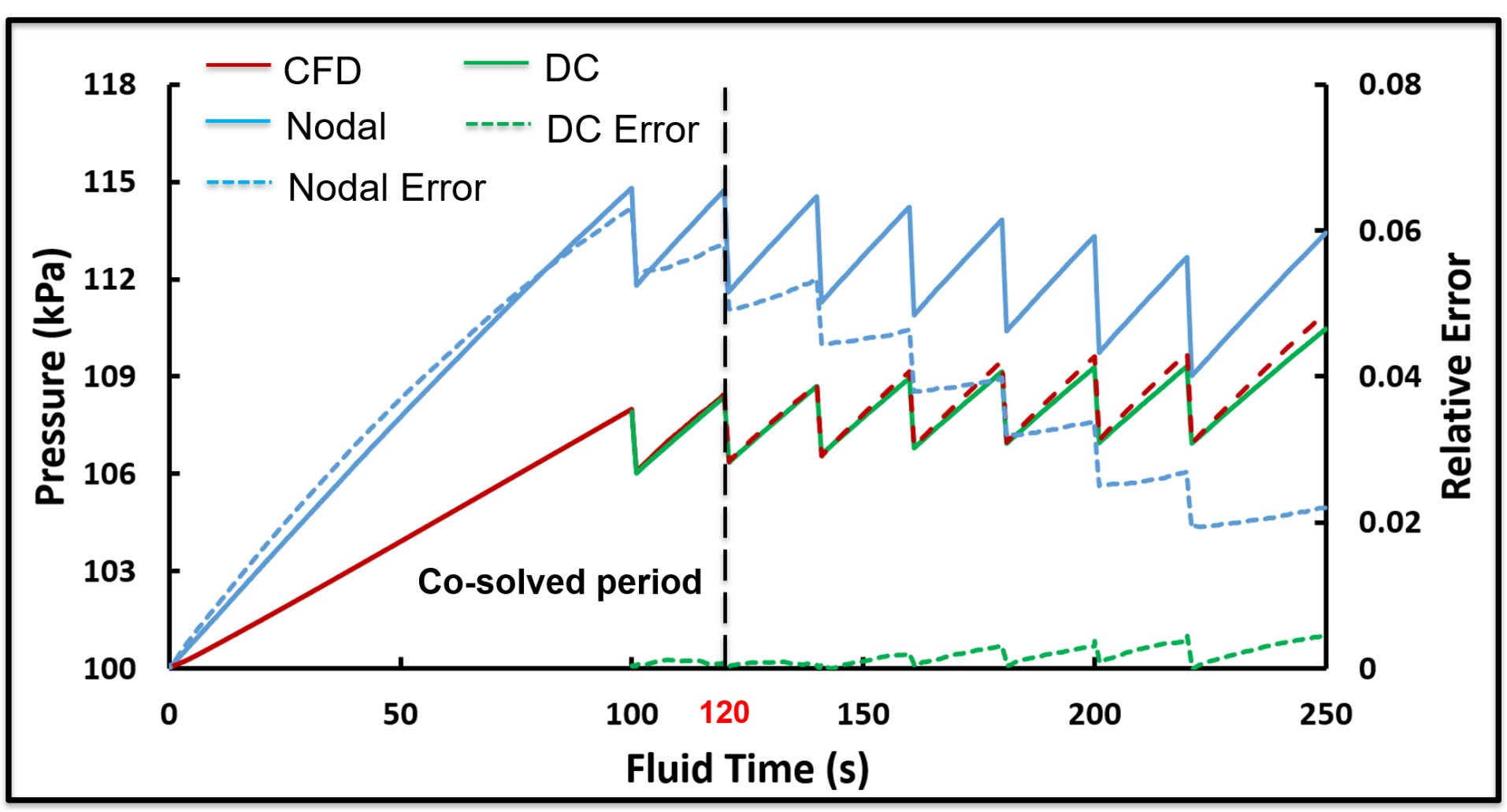

Figure 15. Comparison of the tank pressure for the DCC, CFD and the nodal solutions. The relative errors of the DCC and nodal solutions to the CFD solutions are represented by dashed lines.

## 4.2. Experiment Validation

In this section, self-pressurization experiments are used to validate the performance of the DCC approach. Since the DCC approach is designed to enhance the nodal simulation accuracy through short period co-solving CFD simulations, we employ validated CFD results from the literature as co-solving data and compare the DCC solutions against experimental measurements to assess its performance. Due to the limitations of the simplified nodal model and the availability of experimental data, certain scenarios, such as the periodically operated case in the Thermodynamic Venting System (TVS) tests, are not included in this validation.

### 4.2.1. Multipurpose Hydrogen Test Bed Self-Pressurization Experiments

The Multipurpose Hydrogen Test Bed (MHTB) is a large-scale experimental facility designed to study cryogenic propellant storage and management under low Earth orbit (LEO) conditions at the Marshall Space Flight Center [48]. The MHTB tank is cylindrical in shape, with 3.05 $m$ diameter and 3.05 $m$ height and two 2:1 elliptical end caps. It has an internal volume of 18.09 $m^3$ and a surface area of 35.74 $m^2$, representation of a full-scale liquid hydrogen ($LH_2$) tank. Multiple experiments, including self-pressurization studies and spray bar thermodynamic vent system tests [48], have been conducted using the MHTB. Additionally, CFD models [23] and system-level models [14, 46] have been developed and validated against experimental data from MHTB experiments.

In this section, experimental data from the 50% fill level self-pressurization experiment is used to validate the DCC approach. The MHTB tank is filled with hydrogen to 50% fill level. The initial pressure is 111.6 kPa, and the tank reaches saturation temperature at the start of the experiment. According to a previous study [23], the heat flux is 0.89873 $W/m^2$ in the vapor region and 2.0841 $W/m^2$ in the liquid region. Kartuzova et al. [23] developed a CFD model of the MHTB tank, which was validated against corresponding experimental data. Therefore, we utilize Kartuzova et al.'s CFD simulation data as co-solved data to fit the data-driven correlations. The nodal model of the MHTB tank consists 3 lumps: a vapor lump, a liquid lump, and a twinned lump representing the interface region. The totally simulated flow time is 850 minutes. Due to limitations in available experimental data, only the pressure evolution is analyzed.

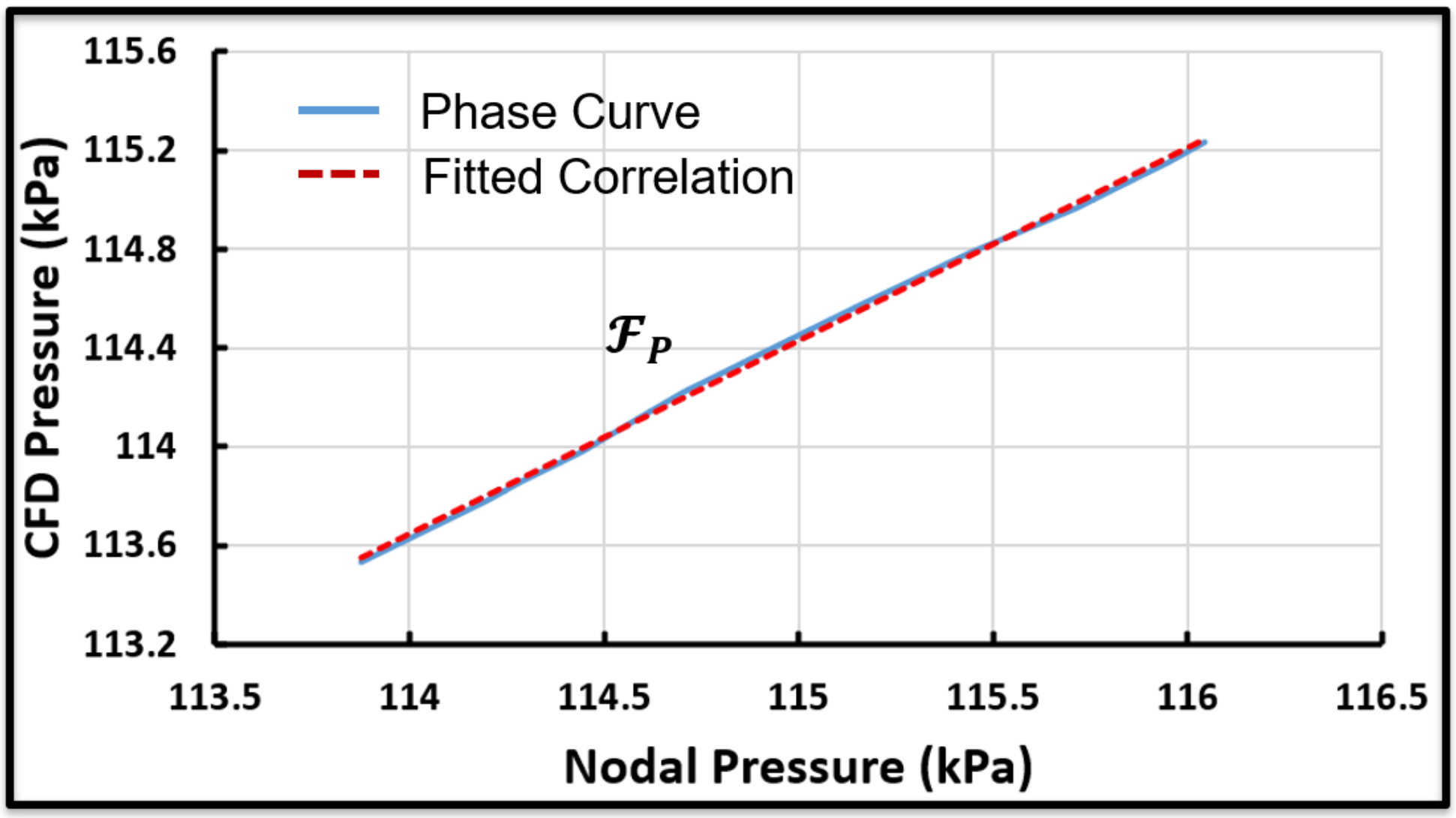

Figure 16. Phase curve of the tank pressure and the fitted correlation ($\mathcal{F}_P$) using the co-solved data from 50 to 100 minutes.

As shown in Figure 16, the DCC approach uses the CFD and nodal simulation results from 50 to 100 minutes as co-solve data. A data-driven correlation for tank pressure is then derived to correct the pressure predictions of the nodal code simulations. Figure 17 compares the tank pressure solutions obtained using the DCC approach against those from CFD simulations, nodal simulations, and experimental data. After the 100-minute co-solving period, the DCC approach extrapolates the data-driven correlation to 850 minutes, which is more than seven times longer than the so-solving period, significantly enhancing the accuracy of the nodal code simulations. To assess the performance of the DCC approach, errors measured in 2-norm between the original nodal simulations and experiment data, as well as between the DCC solutions and experiment data, are calculated. Over the extended extrapolation period from 100 to 850 minutes, the DCC approach achieves an 88% reduction in the 2-norm error. Additionally, at t=800 minutes (the last available experiment data point), the DCC approach reduces the relative error from 6% to 0.8%.

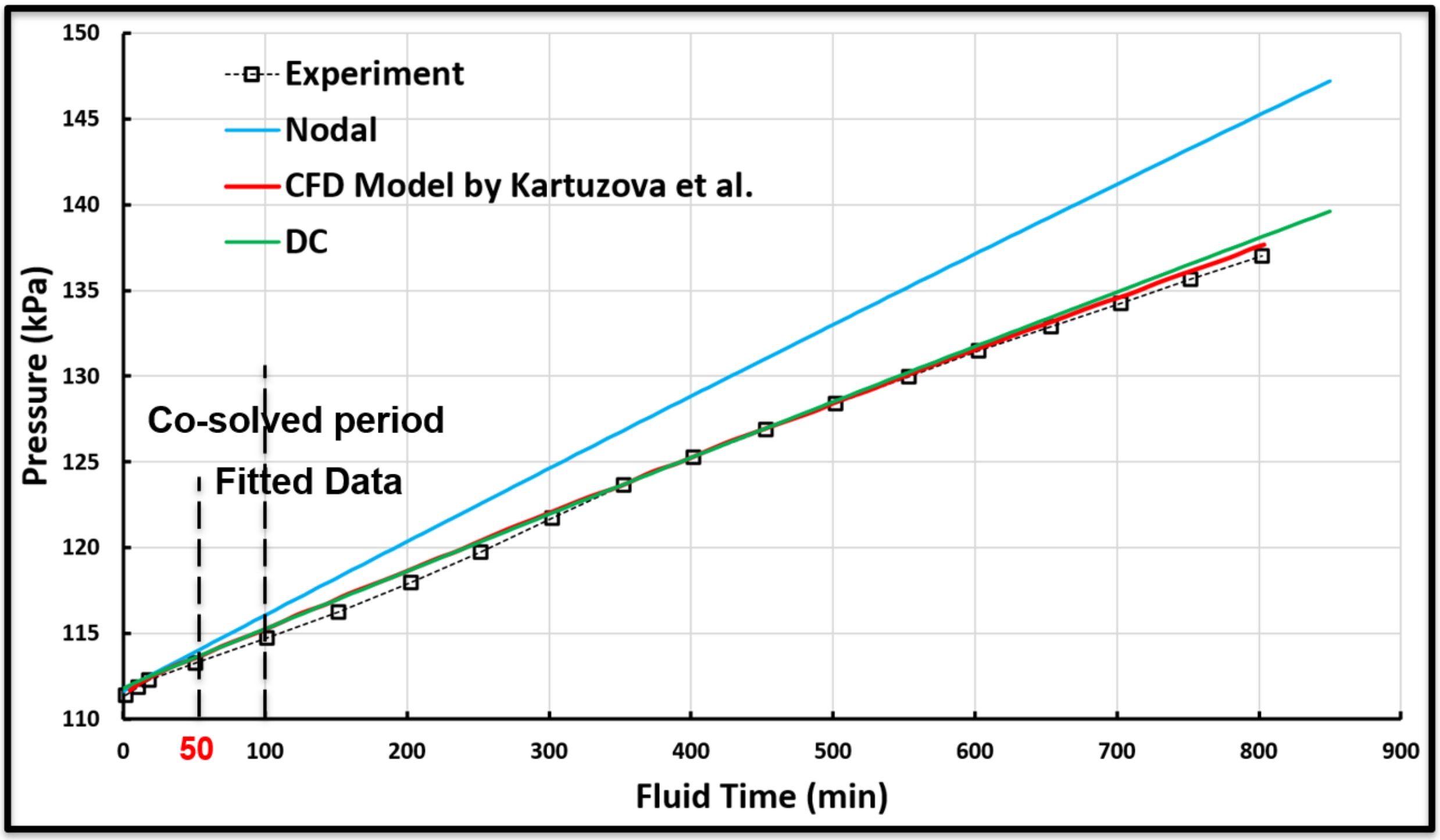

Figure 17. Comparison of the MHTB tank pressure for the DCC, CFD and the nodal solutions.

#### 4.2.2. K-Site Tank Self-Pressurization Experiments

The K-Site tank, developed at NASA's Glenn Research Center (formerly Lewis Research Center), is a flight-weight, cryogenic liquid hydrogen storage tank designed for experimental research. The tank is approximately an ellipsoidal volume of revolution, with a major to minor axis ratio of 1.2, a major diameter of 2.2 $m$, and a volume of 4.89 $m^3$. A series of self-pressurization experiments were conducted using the K-Site tank with various fill levels and applied heat fluxes [53]. In this analysis, we focus on the case with a 50% fill level and an applied heat flux of 3.5 $W/m^2$ to demonstrate the DCC approach. The tank's nodal model includes three lumps: a vapor region, a liquid region, and a twinned lump representing the interface region. For simplification, the tank wall is assumed to have zero thickness, a modeling choice that may overestimate the pressurization rate according to Kartuzova et al.'s work [54]. The CFD simulation data is obtained from Kassemi et al. 's work [52], which was validated against the corresponding experiment data.

Figure 18 illustrates the phase curve of the tank pressure using the CFD and nodal simulation data from 3000 to 6000 seconds (50 to 100 minutes). A data-driven correlation was then derived to map the nodal simulation pressures to the CFD solutions. This correlation was subsequently applied to correct the nodal simulation results up to 65,000 seconds, and the results are compared in Figure 19. Despite the aggressive extrapolation, which extended the correlation nine times beyond the initial timeframe, the DCC approach still demonstrates improved accuracy in an efficient manner. Compared with the original nodal simulation solutions, the DCC approach reduces the error measured in 2-norm by 93.6%, and reduces the relative error at t=61,210 seconds (the last available experiment data point) from 35.8% to 3.0%.

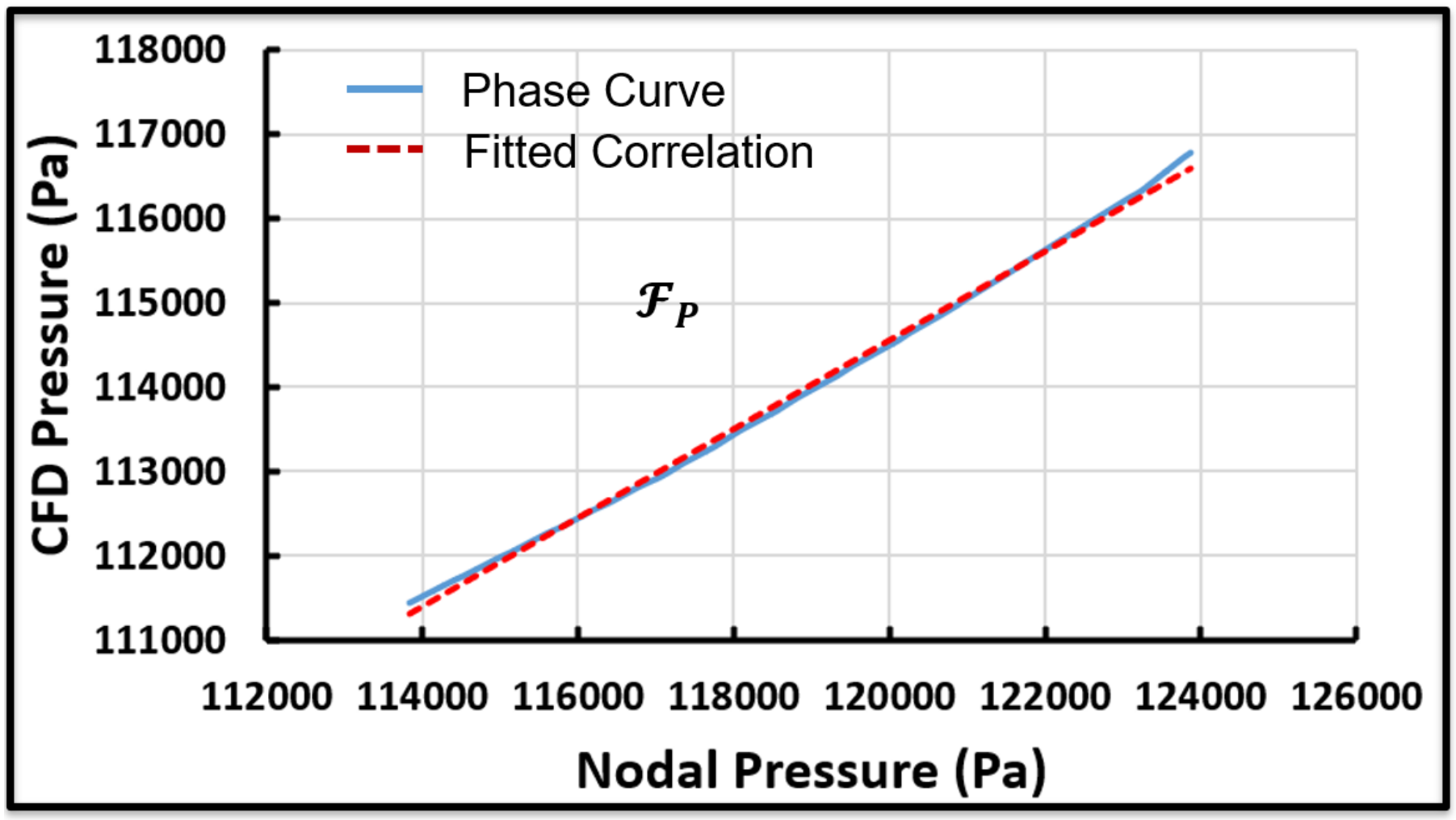

Figure 18. Phase curve of the tank pressure and the fitted correlation ($\mathcal{F}_P$) using the co-solved data from 3000 to 6000 seconds.

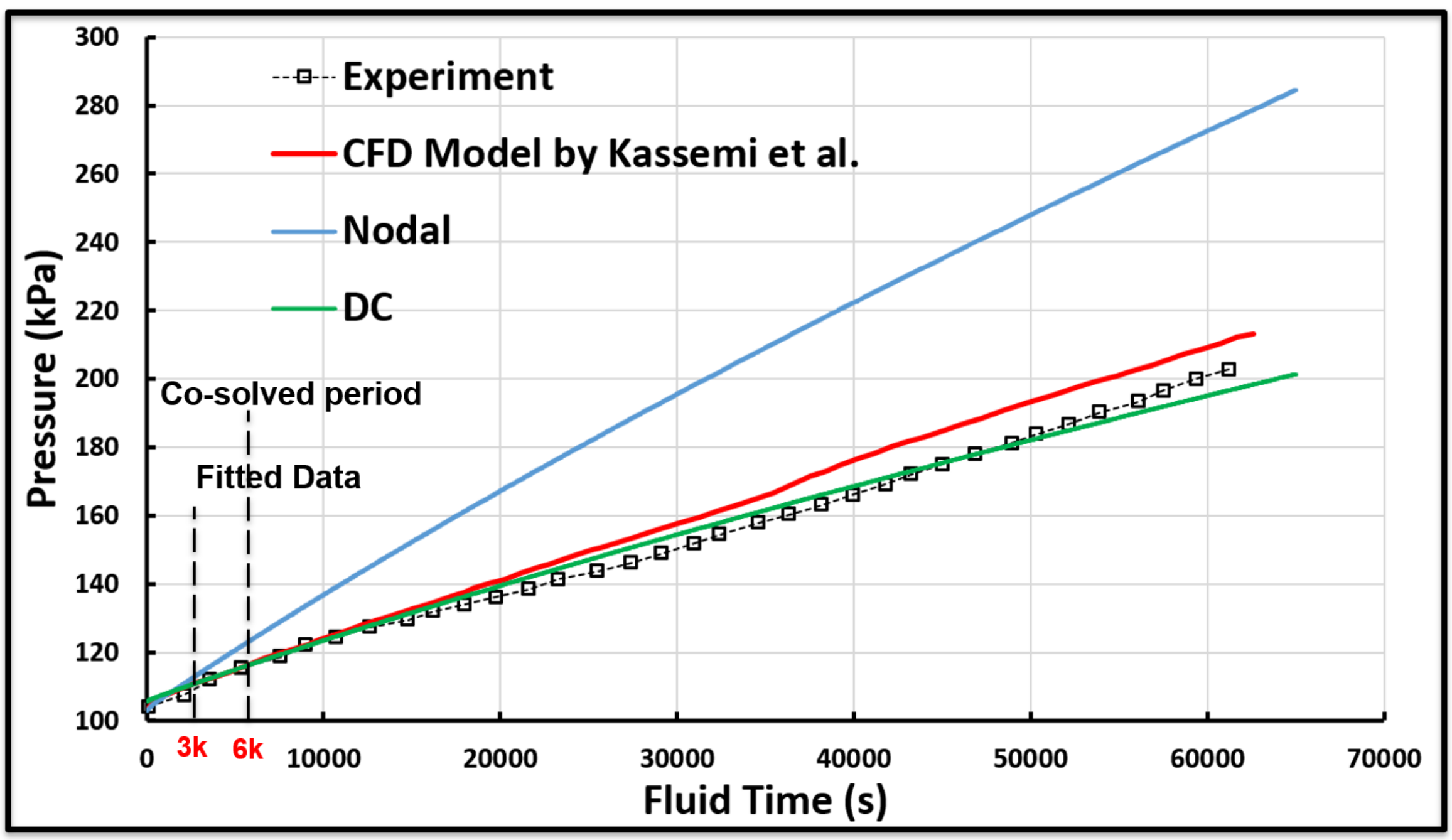

Figure 19. Comparison of the K-site tank pressure for the DCC, CFD and the nodal solutions.

The validation results in Section 4.2.1 and 4.2.2 demonstrate that, provided the CFD simulations are sufficiently accurate, the DCC approach efficiently leverages a short period of co-solved CFD data to enhance the accuracy of nodal simulations over an extended period. Therefore, instead of running time-consuming CFD simulations for the entire problem, the DCC approach reduces the CFD workload to a short co-solving period, improving the overall efficiency of cryogenic tank simulations.

## 5. Conclusion and Discussion

A data-driven based concurrent coupling approach for integrated CFD and nodal analysis has been developed, analyzed, and demonstrated as an effective approach for long-term cryogenic tank pressure control simulations. This approach incorporates data-driven techniques into the concurrent coupling scheme and dynamically generate data-driven correlations from the high-fidelity CFD solutions during a short co-solve period. The obtained correlations bridge the gap between the simplified 1D nodal model and the detailed 3D CFD simulations by mapping the nodal governing equations to the volume-averaged CFD equations. The nodal codes then run exclusively with the data-driven correlations to solve the problem with an improved accuracy.

The DCC approach is demonstrated through both synthetic tests and experiment validations. Synthetic tests include self-pressurization and periodic mixing in a two-phase cryogenic tank. Compared to CFD simulations, the DCC approach significantly reduces computational cost while maintaining comparable accuracy in predicting tank pressure and temperature. Experiment validations are conducted using self-pressurization data from the MHTB and K-Site tanks. For these long-duration, slowly evolving scenarios, the DCC approach achieves notably higher accuracy than nodal simulations by leveraging data-driven correlations derived from a short period of CFD data. These correlations are extrapolated to simulate time periods more than seven and nine times longer than the co-simulation periods for the MHTB and K-Site tanks, respectively. Overall, the DCC approach shows promising potential to support the design and optimization of long-term cryogenic propellant storage tanks for future space missions.

Compared to the spatial domain decomposition coupling approach, the DCC approach minimizes 1D to 3D data transfer, enhancing numerical stability and eliminating the need for iterative solving between CFD and nodal codes within each time step. In contrast to the spatial domain overlapping coupling approach, the DCC approach directly maps the governing equations, avoiding the challenges of designing correction terms and preserving inherent correlations. The data-driven correlations generated through DCC approach are extended over much longer simulation times, rather than being updated in every time step, resulting in a significant boost in computational efficiency. Instead of generating universal empirical correlations, the DCC approach dynamically generates the data-driven correlations by demand, ensuring their applicability throughout the simulation. Furthermore, the DCC approach could leverage real-time measurement data to calibrate fast-execution surrogate models using on-the-fly generated data-driven correlations, making it a potential algorithm for digital twins.

In this work, linear regression was selected to generate the data-driven correlations. Beyond its simplicity and ease of implementation, linear correlations offer the advantage of allowing for an analytical discussion of the physical significance and extrapolation errors associated with the DCC approach. However, linear regression may be insufficient for handling more complex scenarios, and the DCC approach is naturally compatible with advanced data-driven techniques, such as neural networks. A proof-of-concept study demonstrated the potential of a neural network-based method to map 1D nodal results back to 2D CFD solutions [53]. The trained neural network can serve as the data-driven correlation, not only improving the accuracy of nodal solutions but also providing spatial distributions. With these 2D distributions available, after a certain period of extrapolation, the 2D solutions can be transferred back to the CFD solver, initiating a new co-solve period to update or generate new data-driven correlations, thereby enhancing long-term simulation accuracy. This will be the next step for applying the DCC approach to solving more realistic cryogenic propellant tank problems over long-term storage periods in space.

**Acknowledgement**

This work was supported by an Early-Stage Innovations Grant through NASA's Space Technology Research Grants Program under Grant 80NSSC20K0303. The authors would like to thank Dr. Michael F. Harris from NASA Kennedy Space Center for the technical support. The authors would also like to thank the technical support provided by Dr. Mamoru Ishii of Purdue University. Finally, the authors would also like to thank C&R Technologies Inc. (CRTech) for providing the computational software SINDA/FLUINT & Thermal Desktop for this research.

**Appendix. Extrapolations Errors of Polynomial Data-Driven Correlations**

A general polynomial data-driven correlation function $\mathcal{F}\left(T_N^i\right)$ is:

$$\mathcal{F}\left(T_N^i\right) = a_0 + a_1 T_N^i + a_2\left(T_N^i\right)^2 + a_3\left(T_N^i\right)^3 + \cdots. \quad \text{(A.1)}$$

Substitute Equation A.1 and 20 to Equation 23:

$$T_{DCC}^{n+1} = \mathcal{F}(T_N^n + Hr_N^n * \Delta t) = a_0 + a_1(T_N^n + Hr_N^n * \Delta t) + a_2(T_N^n + Hr_N^n * \Delta t)^2 + \cdots, \quad \text{(A.2)}$$

Since $Hr_N^n * \Delta t$ can be treated as a small quantity to the $T_N^n$, higher order terms of $\Delta t$ can be neglected. Rearrange the Equation A.2:

$$\begin{aligned} T_{DCC}^{n+1} &= (a_0 + a_1 T_N^n + a_2(T_N^n)^2 + a_3(T_N^n)^3 + \cdots) \\ &\quad + \left(a_1 Hr_N^n \Delta t + a_2 T_N^n Hr_N^n \Delta t + a_3 {T_N^n}^2 Hr_N^n \Delta t + \cdots\right) \\ &= \mathcal{F}(T_N^n) + \left(a_1 Hr_N^n \Delta t + a_2 T_N^n Hr_N^n \Delta t + a_3 {T_N^n}^2 Hr_N^n \Delta t + \cdots\right) \\ &= T_{DCC}^n + \left(a_1 Hr_N^n \Delta t + a_2 T_N^n Hr_N^n \Delta t + a_3 {T_N^n}^2 Hr_N^n \Delta t + \cdots\right). \end{aligned} \quad \text{(A.3)}$$

Therefore, high order polynomials correlations will introduce $T_N^n \Delta t$ into correction terms, which grows fast especially when the temperature is high. To eliminate the influence of the $T_N^n$ terms, a linear equation is selected as the data-driven correlation function.